\documentclass{aa}  

\usepackage{color}
\usepackage{xcolor}
\definecolor{xlinkcolor}{cmyk}{1,1,0,0}

\usepackage{multicol}

\usepackage{graphicx} 
\usepackage[varg]{txfonts} 
\usepackage[activate={true,nocompatibility},
    final,
    tracking=true,
    kerning=true,
    spacing=true,
    factor=1100,
    stretch=10,
    shrink=10]{microtype}
\usepackage[bookmarks=true,         % show bookmarks bar?
     bookmarksopen=true,
     pdfnewwindow=true,      % links in new window
     colorlinks=true,    % false: boxed links; true: colored links
     linkcolor=xlinkcolor,     % color of internal links
     citecolor=xlinkcolor,     % color of links to bibliography
     filecolor=xlinkcolor,  % color of file links
     urlcolor=xlinkcolor,      % color of external links
     final=true]{hyperref} 
\hypersetup{pdfstartview=Fit}
\usepackage{bookmark}

\usepackage{longtable}

\usepackage{adjustbox}
\allowdisplaybreaks

\usepackage[labelfont=bf]{caption}

\usepackage{etoolbox}
\makeatletter
\patchcmd\@combinedblfloats{\box\@outputbox}{\unvbox\@outputbox}{}{\errmessage{\noexpand patch failed}}
\makeatother

\makeatletter
\renewcommand*\aa@pageof{, page \thepage{} of \pageref*{LastPage}}
\makeatother

\usepackage{afterpage}


\newcommand\Tstrut{\rule{0pt}{2ex}}         % = `top' strut
\newcommand\Bstrut{\rule[-0.8ex]{0pt}{0pt}}   % = `bottom' strut

\makeatletter
\newif\ifhbonecolumn
\@ifclasswith{aa}{referee}{\hbonecolumntrue}{\hbonecolumnfalse}
\makeatother

\newcommand{\thinfig}{\ifhbonecolumn 0.66\linewidth\else 1\linewidth\fi}

\newcommand{\bottrim}{\ifhbonecolumn 1cm\else 1.2cm\fi}
\newcommand{\lefttrim}{\ifhbonecolumn 1.1cm\else 1.3cm\fi}
\newcommand{\righttrim}{\ifhbonecolumn 0.8cm\else 1cm\fi}

\newif\ifref
\reffalse
\newcommand{\mb}[1]{\ifref\boldmath\textbf{#1}\unboldmath\else #1\fi}

\newif\ifreff
\refffalse
\newcommand{\mbb}[1]{\ifreff\boldmath\textbf{#1}\unboldmath\else #1\fi}

\newif\ifrefff
\reffffalse
\newcommand{\mbbb}[1]{\ifrefff\boldmath\textbf{#1}\unboldmath\else #1\fi}

\newcommand{\LEt}[1]{}

\begin{document} 

   %\title{How do stellar radius determinations depend on metallicity?} 
   \title{Stellar ages, masses, and radii from asteroseismic modeling are robust to systematic errors in spectroscopy}
   %\title{Stellar ages from asteroseismology are robust\\to systematic errors in spectroscopy} 
   %\subtitle{}
   \titlerunning{The impact of systematic errors in spectroscopic measurements on stellar parameters} 
   %\authorrunning{E.\ P.\ Bellinger et al.\ 2018}
   %\subtitle{Assessing the impact of systematic errors in [Fe/H] measurements on parameters derived from stellar modelling for solar-like stars} 

   \author{E.~P.~Bellinger\inst{1}\fnmsep\thanks{SAC Postdoctoral Fellow}
          \and
          S.~Hekker\inst{2,1}
          \and
          G.~C.~Angelou\inst{3}
          \and
           A.~Stokholm\inst{1}
          \and
          S.~Basu\inst{4}
          %\and
          %J. Christensen-Dalsgaard\inst{1}
          %\and
          %...
          }

   \institute{Stellar Astrophysics Centre, Department of Physics and Astronomy, Aarhus University, Denmark \\
              \email{\url{bellinger@phys.au.dk}}
         \and
              Stellar Ages \& Galactic Evolution (SAGE) Group, Max Planck Institute for Solar System Research, G\"ottingen, Germany
             %\email{hekker@mps.mpg.de}
         \and 
              Max Planck Institute for Astrophysics, Garching, Germany
         \and
              Department of Astronomy, Yale University, New Haven, Connecticut, USA
         %\and 
         %     ...
             }

   \date{Received 19 October 2018; Accepted 13 December 2018; Published 07 February 2019}

% \abstract{}{}{}{}{} 
% 5 {} token are mandatory

\abstract{The search for twins of the Sun and Earth relies on accurate characterization of stellar and the exoplanetary parameters age, mass, and radius. In the modern era of asteroseismology, parameters of solar-like stars are derived by fitting theoretical models to observational data, which include measurements of their oscillation frequencies, metallicity [Fe/H], and effective temperature $T_{\text{eff}}$. Furthermore, combining this information with transit data yields the corresponding parameters for their associated exoplanets.} {While values of [Fe/H] and $T_{\text{eff}}$ are commonly stated to a precision of $\sim$~0.1~dex and $\sim$~100~K, the impact of systematic errors in their measurement \mbbb{has} not been studied in practice within the context of the parameters derived from them. Here we seek to quantify this.} {We used the \emph{Stellar Parameters in an Instant} (SPI) pipeline to estimate the parameters of nearly 100 stars observed by \emph{Kepler} and \emph{Gaia}, many of which are confirmed planet hosts. We adjusted the reported spectroscopic measurements of these stars by introducing faux systematic errors and, \mb{separately,} artificially increasing the reported uncertainties of the measurements\mbbb{, and quantified} the differences in the resulting parameters\LEt{Please check I have retained your intended meaning. }.} {We find that a systematic error of 0.1~dex in [Fe/H] translates to differences of only 4\%, 2\%, and 1\% on average in the resulting stellar ages, masses, and radii, which are well within their uncertainties ($\sim$~11\%, 3.5\%, 1.4\%) \mbb{as derived by SPI}. We also find that increasing the uncertainty of [Fe/H] measurements by 0.1~dex increases the uncertainties of the ages, masses, and radii by only 0.01~Gyr, 0.02~$\text{M}_\odot$, and 0.01~$\text{R}_\odot$, which are again well below their reported uncertainties ($\sim$~0.5~Gyr, 0.04~$\text{M}_\odot$, 0.02~$\text{R}_\odot$). The results for $T_{\text{eff}}$ at 100~K are similar.} {\mb{Stellar parameters from SPI are unchanged within uncertainties by errors of up to 0.14~dex or 175~K. They are even more robust to errors in $T_{\text{eff}}$ than the seismic scaling relations. Consequently, the parameters for their exoplanets are also robust.}}
 
    
   \maketitle
%
%________________________________________________________________

\section{Introduction} 

The modern study of cool dwarf stars has been revolutionized in recent years by ultraprecise measurements of low-amplitude global stellar pulsations (for a summary of early results, see \citealt{2013ARA&A..51..353C}). 
Traveling through the star before emerging near the surface, pulsations bring information to light about the otherwise opaque conditions of the stellar interior, and thereby provide numerous constraints to the internal structure and composition of stars. 
Indeed, connecting theoretical stellar models to measurements of pulsation frequencies---as well as other measurements, such as those derived from spectroscopy---yields precise determinations of stellar parameters such as age, mass, and radius \citep[for an overview, see][]{basuchaplin2017}. 

Precisely constrained stellar parameters are broadly useful for a variety of endeavours, such as testing theories of stellar and galactic evolution \citep[e.g.,][]{2016A&A...589A..93D, 2017MNRAS.464.3713H, 2017A&A...608A.112N, 2017ApJ...851...80B} and mapping out history and dynamics of the Galaxy \citep[e.g.,][]{2015A&A...576L..12C, 2017A&A...597A..30A, 2018MNRAS.475.5487S, 2018MNRAS.477.2326F}. 
Combining stellar parameters with transit measurements furthermore yields properties of the exoplanets that they harbor, such as their masses, radii, obliquities, eccentricities, and semi-major axes \mbbb{\citep[e.g.,][]{2003ApJ...585.1038S, 2013Sci...342..331H, 2014ApJ...782...14V, 2016IAUFM..29B.620H, 2018ASSP...49..119H, 2016ApJ...819...85C, 2018MNRAS.479..391K, 2018ASSP...49..225A}}. 
This then permits tests to theories of planet formation \citep[e.g.,][]{1996Natur.380..606L, 2011MNRAS.413L..71W, 2011ApJ...729..138M, 2011MNRAS.412.2790L, 2011ApJ...732...74G, 2011MNRAS.417.1817T, 2013MNRAS.428..658T, 2013ApJ...769...86P, 2015ApJ...809L..20M, 2017AJ....153...60M, 2016ApJ...818....5L, 2017ApJ...846L..13K, 2017MNRAS.464.1709G} and facilitates the search for habitable exoplanets.

%\subsection{Sources of systematic errors in spectroscopy}

%\subsection{Asteroseismic modelling}

From first principles, one may obtain approximate relations to deduce stellar radii ($R$), mean densities ($\rho$), and masses ($M$) from asteroseismic data by scaling the values of the Sun to the observed properties of other stars \citep[``scaling relations,''][]{1986ApJ...306L..37U, 1991ApJ...368..599B, 1995A&A...293...87K}:
%\vspace*{-0.1cm}
\begin{align}
%\begin{align} 
%\end{equation}
%\begin{equation}
    \frac{R}{\text{R}_\odot}
    %\simeq
    &\simeq
    \left(
        \frac{\nu_{\max}}{\nu_{\max,\odot}}
    \right)
    \left(
        \frac{\Delta\nu}{\Delta\nu_\odot}
    \right)^{-2}
    \left(
        \frac{T_{\text{eff}}}{T_{\text{eff},\odot}}
    \right)^\frac{1}{2} \label{eq:scalingR}
\end{align}\clearpage\vspace*{-2.7\baselineskip}%\ifref\else\fi
\begin{align}
%    \\
    \frac{\rho}{\rho_\odot}
    %\simeq%
    &\simeq
    \left(
        \frac{\Delta\nu}{\Delta\nu_\odot}
    \right)^{2} \label{eq:scalingRho}
    \\
    \frac{M}{\text{M}_\odot}
    %\simeq %
    &\simeq
    \left(
        \frac{\nu_{\max}}{\nu_{\max,\odot}}
    \right)^3
    \left(
        \frac{\Delta\nu}{\Delta\nu_\odot}
    \right)^{-4}
    \left(
        \frac{T_{\text{eff}}}{T_{\text{eff},\odot}}
    \right)^\frac{3}{2}. \label{eq:scalingM}
%\end{equation}
%\begin{equation}
%\end{equation}
\end{align}
Here $T_\text{eff}$ is the effective temperature of the star, $\nu_{\max}$ is the frequency of maximum oscillation power, and $\Delta\nu$ is the large frequency separation \citep[for detailed definitions, see, e.g.,][]{basuchaplin2017}. 
The quantities subscripted with the solar symbol ($\odot$) correspond to the solar values: ${\nu_{\max,\odot} = 3090\pm30~\mu\text{Hz}}$, ${\Delta\nu_\odot = 135.1\pm 0.1~\mu\text{Hz}}$, and ${T_{\text{eff},\odot} = 5772.0\pm0.8~\text{K}}$ \citep{2011ApJ...743..143H, 2016AJ....152...41P}. 
This way of estimating stellar parameters is often referred to as the ``direct method'' \citep[e.g.,][]{2018arXiv180402214L} \mb{because it is independent of stellar models and relies only on physical arguments}. 
We note that there are no such scaling relations for the stellar age. 
%Obviously, a combination of these relations also yields the stellar mean density ($\rho_\ast$). 

% The scaling relations have been applied to estimate stellar parameters for more than 500 main sequence and subgiant stars (Chaplin et al. 2014) and thousands of red giants (e.g., Pinsonneault et al. 2018; Silva Aguirre et al. 2018) observed by the Kepler spacecraft .

The mass and radius seismic scaling relations are functions of the stellar effective temperature, which, along with metallicity, can be derived from spectroscopic observations of a star. 
%Frequently the question of their reliability has been raised. 
A variety of factors contribute \mbbb{to} the determination of these spectroscopic parameters, such as spectra normalization, corrections for pixel-to-pixel variations, the placement of the continuum, and so on  \citep[e.g.,][]{2013pss2.book...35M, 2011arXiv1112.2787S}. 
The analysis can be complicated by effects such as line blending and broadening. 
Uncertainties in atmospheric models and the techniques for obtaining atmospheric parameters such as the measurement of equivalent widths, fitting on synthetic spectra, and degeneracies between spectroscopic parameters can also cause difficulties. 
%The analysis sometimes relies on visual inspection, and so different practitioners can come to different conclusions about the same star. 
Often, parameters for targets in large spectroscopic surveys are extracted automatically without visual inspection, which can lead to poor fits that sometimes go unnoticed. 
Conversely, analysis based on visual inspection sometimes leads to different practitioners coming to different conclusions about the same star. 
Thus there are opportunities for systematic errors to be introduced into the determined effective temperatures and metallicities \mb{\citep[e.g.,][]{2013MNRAS.431.2419C}}. %, and the question of their reliability has frequently been raised. 
However, to date, no study has \mb{systematically} tested in practice the impact of systematic errors in spectroscopic parameters on the determination of stellar parameters \mb{for a large sample of stars}. 
% \citep[e.g.,][]{2015ApJ...811L..37M}. 

Unlike the effective temperature, the classical scaling relations presented above are not functions of metallicity, and so one would expect metallicity to have little importance in the determination of stellar masses and radii. 
However, these relations are approximations, and are not perfectly accurate. 
Several studies \citep[e.g.,][]{2011ApJ...743..161W, 2011A&A...530A.142B, 2012ApJ...749..152M, 2013A&A...556A..59H, 2013A&A...550A.126M, 2014ApJS..211....2H, 2016MNRAS.460.4277G, 2016ApJ...822...15S, 2016ApJ...832..121G, 2017ApJ...844..102H, 2017MNRAS.470.2069G, 2017ApJ...843...11V, 2018MNRAS.478.4669T} have sought to quantify the validity of the seismic scaling relations---especially as a function of stellar evolution, when the assumption of solar homology breaks down---for example by comparison with stellar models, or via orbital analyses of binary stars. 
\citet{2017ApJ...843...11V} have also recently shown theoretically that these relations neglect a term for the mean molecular weight, which implies a previously unaccounted for dependence on metallicity. 

To obtain more accurate mass and radius estimates---and stellar ages---one can fit theoretical models to the observations of a star \citep[e.g.,][]{1994ApJ...427.1013B,  2011ApJ...730...63G, 2014ApJS..214...27M,  2014A&A...569A..21L, 2015MNRAS.452.2127S,  2017ApJ...835..173S, 2016ApJ...830...31B}. 
The connection between observations and stellar parameters then becomes much more complex, relying on the details of not only the coupled differential equations governing the structure and evolution of stars, but also the choices and implementations regarding the microphysics of the star, such as the opacities and equation of state of stellar plasma. %, and the rates at which the various nuclear reactions occur. %nuclear reaction rates. 
This motivates\LEt{replace with carifies?} the need for numerical studies to determine the relationships between the observations of a star and the resulting fundamental parameters \citep[e.g.,][]{1994ApJ...427.1013B, 2017ApJ...839..116A, 2018arXiv181006997V}. 
Here we examine one important aspect of these relations: namely, the impact of systematic errors and underestimated uncertainties in spectroscopic parameters on the determination of stellar parameters via asteroseismic modeling. 

In order to perform this study, we used the \emph{Stellar Parameters in an Instant} pipeline \citep[SPI,][]{2016ApJ...830...31B}, which uses machine learning to rapidly connect observations of stars to theoretical models. 
\mb{The SPI method involves training an ensemble or ``forest'' of decision tree regressors \citep{breiman2001random, geurts2006extremely, friedman2001elements} to learn the relationships between the observable aspects of stars (such as their oscillations) and the unobserved or unobservable aspects (such as their age). 
We then fed the observations of a particular star into the learned forest, which outputs the desired predicted quantities. 
In order to propagate the uncertainties of the observations into each of the predicted quantities, we generated random realizations of noise according to the \mbb{uncertainties of the observations (including their correlations)}, and ran these random realizations also through the forest. 
This yields the posterior distribution for each predicted quantity, from which we may obtain, for example, the mean value and its corresponding uncertainty for quantities such as the stellar age, mass, radius, mean density, and luminosity.
}
%This yields posterior distributions for each predicted quantity, and obtain posterior distributions (and therefore uncertainties) fo }
%From this posterior distribution we may obtain, for example, a mean and uncertainty for, e.g., the predicted age, mass, radius, luminosity, and density. } 

Using this approach, one can simultaneously and independently vary multiple free parameters corresponding to uncertain aspects of stellar models, such as the efficiency of convection \mb{and the strength of gravitational settling}; and therefore propagate many of the uncertainties of the theoretical models into the resulting stellar parameters. 
\mb{To this end, we generated a large grid of stellar models that have been varied in age as well as seven other parameters controlling the physics of the evolution, which is described in more detail in the next section. } 
Since obtaining the parameters of a star with SPI takes less than a minute (rather than hours or days, as is typical with some other methods) we were then able to test the robustness of the final stellar parameters against injections of various amounts of systematic errors. 
%These include the strength of gravitational settling and diffusion, the extent of convective core overshooting and convective envelope undershooting, and the efficiency of convection. 

\LEt{Subsections are not allowed in the Introduction.}

\begin{figure}
    \centering
    \makebox[\thinfig][c]{%
        \adjustbox{trim=0cm 0cm 0cm 0.44cm, clip}{%
            \includegraphics[width=\thinfig]{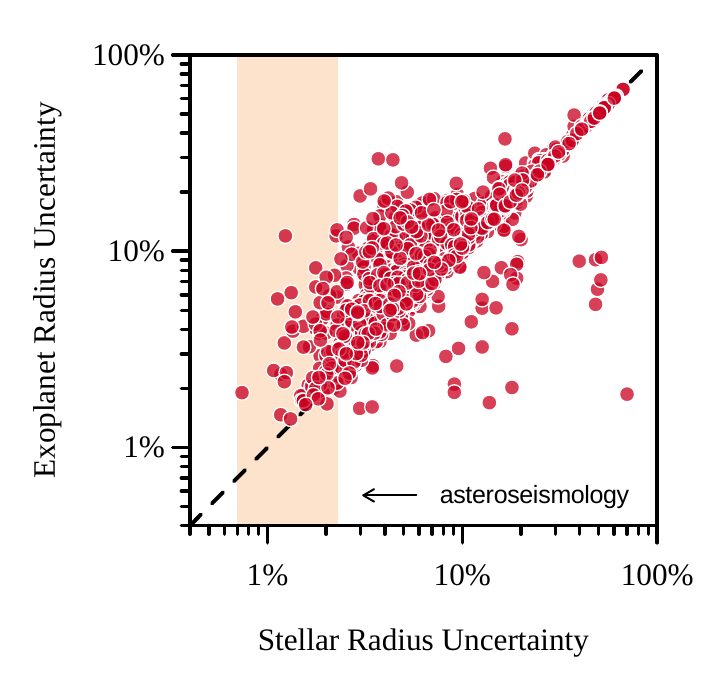}
    }}
    \caption{Relative uncertainties of the radii of transiting exoplanets as a function of the relative uncertainty in the radius of their host star. 
    The dashed line shows a one-to-one agreement. % and the dotted line is the line of best fit. 
    The shaded region shows the full range of stellar radii uncertainties determined for the 97 stars examined in this paper. 
    Exoplanet data are taken from the collection by \citet{2014PASP..126..827H}.
    } 
    \label{fig:exo-R}
\end{figure}

%Beyond the immediate implications, 
The question of how systematic errors in spectroscopy impact stellar parameter determinations has direct bearing on exoplanet research. 
This is due to the fact that the determination of exoplanetary parameters generally depends very strongly on the ability to constrain the parameters of the host star (see Figure~\ref{fig:exo-R}). 

\begin{figure}
    \centering
    \makebox[\thinfig][c]{%
        %\adjustbox{trim=0cm 0cm 0cm 0.33cm, clip}{%
        \adjustbox{trim=0cm 0cm 0cm 0.44cm, clip}{%
            \includegraphics[width=\thinfig]{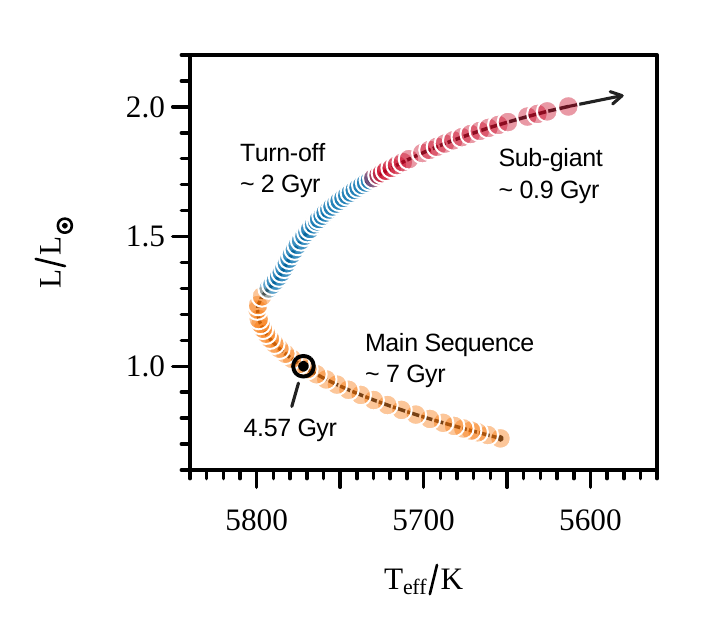}
        }%
    }%
    \caption{Theoretical evolution of the Sun in the HR diagram. 
        The position of the present-age Sun is given with the solar symbol ($\odot$). 
        The points indicate models that have been selected from the track. 
        They are colored according to their assigned phase of evolution, and the approximate duration of each phase is indicated. 
        %The colors of the points correspond to the indicated phases of evolution, and the duration of those phases are indicated. 
    \label{fig:solar-ev}}
\end{figure}

Given the radius of the host star from asteroseismology (e.g., using Equation~\ref{eq:scalingR}; or using more sophisticated methods, such as SPI), and assuming a uniformly bright stellar disk, the radius of a transiting exoplanet can be found by measuring the transit depth \citep[e.g.,][]{2003ApJ...585.1038S}.  %\citep[e.g.,][]{2013ApJS..204...24B, 2013ApJ...767..127H, 2016ApJ...816...95G}. 
Given the stellar radius and mean density (e.g., Equation~\ref{eq:scalingRho}), and assuming a circular orbit, one can compute the semi-major axis of an exoplanet from its orbital period. 
Combining that information with the stellar mass (e.g., Equation~\ref{eq:scalingM}), one can find the mass of an exoplanet using Kepler's third law. % from its orbital period and semi-major axis
As prevailing theories give that planets form roughly around the same time as the host star, the stellar age that is found is generally then attributed to its companions as well \citep[e.g.,][]{2015AJ....150...56J, 2015ApJ...799..170C, 2016IJAsB..15...93S, 2018arXiv180303125C}. %\citep[e.g.,][]{2015ApJ...799..170C}. 
From this information, the composition of an exoplanet can be determined by connecting these properties to theoretical models of planet structure and formation \citep[e.g.,][]{2007ApJ...669.1279S, 2014ApJ...783L...6W, 2015ApJ...801...41R, 2018ASSP...49..119H}. 

Thus, if the parameters derived for the host star are biased, then so too will be the parameters for its exoplanets. 
A differential bias---e.g., a bias that affects mass more than radius---would furthermore impact strongly on matters such as empirical mass--radius relations \citep[e.g.,][]{2007ApJ...669.1279S, 2010A&ARv..18...67T}.

\section{Methods} \label{sec:methods}

We used \emph{Modules for Experiments in Stellar Astrophysics} \citep[MESA~r10108,][]{2011ApJS..192....3P, 2013ApJS..208....4P, 2015ApJS..220...15P, 2018ApJS..234...34P} to construct a grid of stellar models following the procedure given by \citet[][]{2016ApJ...830...31B}. 
The initial parameters were varied quasi-randomly in the ranges given in Table~\ref{tab:ranges}. 
We introduced an additional free parameter, $\alpha_{uv}$, which controls the efficiency of convective envelope undershooting. 
%We used the classical `step' treatment of overshooting, which extends the convection zone beyond the Schwarzschild boundary by a distance of ${\alpha H_p}$, where $H_p$ is the local pressure scale height and $\alpha$ is either the overshooting ($\alpha_{\text{ov}}$) or undershooting ($\alpha_{\text{us}}$) parameter, depending on the type of convection zone. 
As convection is modeled in MESA as a time-dependent diffusive process, under- and over-shooting are achieved by applying convective velocities to zones within a distance of ${\alpha H_p}$ beyond the convective boundary, where $\alpha$ is the under- or over-shooting parameter, and $H_p$ is the local pressure scale height. 
The convective velocities that are used are taken from a distance $f_0$ before the boundary; here we used ${f_0 = 0.01~H_p}$. 
The remaining aspects of the models are the same as in \citealt{2016ApJ...830...31B}.

\begin{table}%[b]
    \centering%\captionsetup{width=\captwidth}%,font=small}
    %\captionsetup{justification=centering}
%\begin{supertabular}[c]{lcccc}
    %\begin{minipage}{0.5\textwidth}
    %\begin{center}
    \caption{Parameter ranges for the grid of stellar models. 
        For reference, values corresponding to a solar-calibrated model and also the derived initial spectroscopic parameter ranges are listed as well. 
        \label{tab:ranges}} 
    %\end{center}
    %\end{minipage}
%\multicolumn{1}{l}{Parameter} & 
%\multicolumn{1}{c}{Symbol} & 
%\multicolumn{1}{c}{Unit} & 
%\multicolumn{1}{c}{Range} & 
%\multicolumn{1}{c}{Solar Value} \\\hline\hline
    %\caption{Parameter ranges for the grid of stellar models. \label{tab:ranges}}
    \begin{tabular}{lcccc}%\hline\hline
        Parameter & Symbol & Range & Solar value \Bstrut{} \\\hline\hline
        Mass      & $M$/M$_\odot$ & $(0.7, 1.8)$ & 1 \Tstrut{} \\
Mixing length     & $\alpha_{\text{MLT}}$  & $(1, 3)$ & 1.85 \\
Initial helium & $Y_0$  & $(0.22, 0.34)$ & 0.273 \\
Initial metallicity      & $Z_0$  & $(0.0001, 0.04)^{a}$  & 0.019 \\
Overshoot                & $\alpha_{\text{ov}}$  & $(0.0001, 1)^{a}$ & - \\
Undershoot               & $\alpha_{\text{us}}$  & $(0.0001, 1)^{a}$ & $\leq 0.05^{b}$ \\ 
Diffusion factor         & $D$ & $(0.0001, 3)^{a}$ & 1 \Bstrut{}\\\hline%\hline
Eff.\ temperature        & $T_{\text{eff}}$ & $(4000, 14000)$ & 5772 \Tstrut{}\\
Metallicity              & [Fe/H] & $(-2.2, 0.44)$ & 0 \Bstrut{} \\\hline
    %\multicolumn{5}{l}{\footnotesize{$^a$varied logarithmically, $^b$\citealt{1997MNRAS.288..572B}}}%\\%\hline
    \end{tabular}\\[0.2\baselineskip]%\hline
    \footnotesize{$^a$varied logarithmically, $^b$\citealt{1997MNRAS.288..572B}}%\\
    %\hline
%\end{supertabular}
\end{table}

We calculated ${N = 8\,170}$ evolutionary tracks which we simulated from the zero-age main sequence until either an age of $20~\text{Gyr}$ or an asymptotic period spacing of $150~\text{s}$, which is generally around the base of the red giant branch. 
\mb{Since the initial conditions of the grid are varied quasi-randomly, the resolution in each parameter $x$ is given by ${(x_{\max} - x_{\min})/N}$; e.g., the typical resolution in mass is approximately $0.0001~{\text{M}}_\odot$. } 
From each track we obtained 32 models with core hydrogen abundance $X_c > 0.1$ nearly evenly spaced in age (i.e., main-sequence models), 32 models with $0.1 \leq X_c \leq 10^{-6}$ nearly evenly spaced in $\log X_c$ (turn-off models), and 32 models beyond $X_c < 10^{-6}$ again nearly evenly spaced in age (sub-giant models). 
These phases are visualized in Figure~\ref{fig:solar-ev}.

We calculated $p$-mode frequencies for these models using the GYRE oscillation code \citep[][]{2013MNRAS.435.3406T}. 
%We removed mixed oscillation modes in the observable region, which we estimated to be ${\nu_{\max} \pm 7.5\Delta\nu}$. 
We obtained frequency separations and ratios \citep{2003A&A...411..215R} for these models following the procedures given in \citealt{2016ApJ...830...31B}. 
We removed models where the presence of mixed modes made it impossible to calculate these quantities. 
%We processed the theoretical mode frequencies in the same way that we processed the observational data. 

\section{Data} 
\begin{figure}
    \centering
    \makebox[\thinfig][c]{%
        \adjustbox{trim=0cm 0cm 0cm 0cm, clip}{%
            \includegraphics[width=\thinfig]{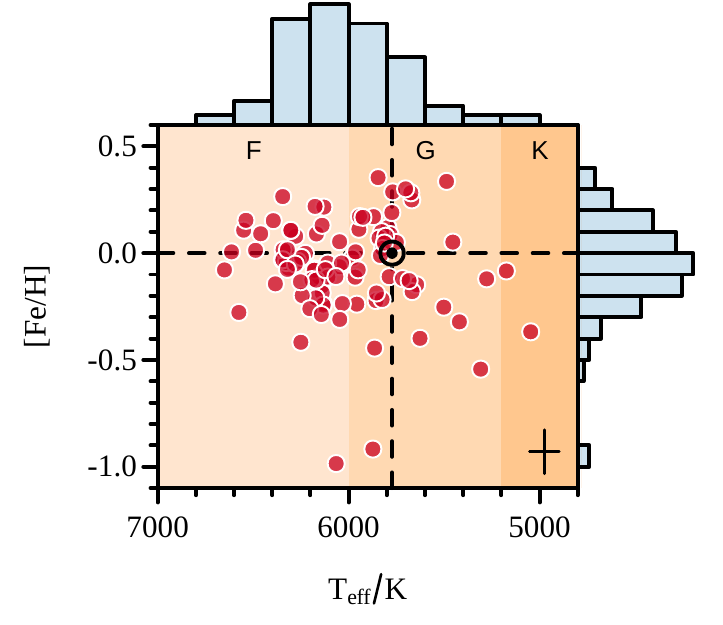}
        }%
    }%
    \caption{Spectroscopic measurements of 97 solar-like oscillating stars observed by \emph{Kepler}. 
    Typical uncertainties (0.1~dex, 100~K) are indicated with the cross in the bottom right corner. 
    Histograms are affixed to the top and right side of the figure showing the metallicity and effective temperature distributions for the sample. 
    The values for the Sun are given with the solar symbol ($\odot$). 
    The background colors indicate the spectral type (F, G, K). 
        \label{fig:spectroscopy}}
\end{figure}

\begin{figure*}%[p]
    \centering
    \makebox[\linewidth][c]{%
        \adjustbox{trim=0cm {\bottrim} 0cm 0cm, clip}{%
            \includegraphics[width=0.5\linewidth]{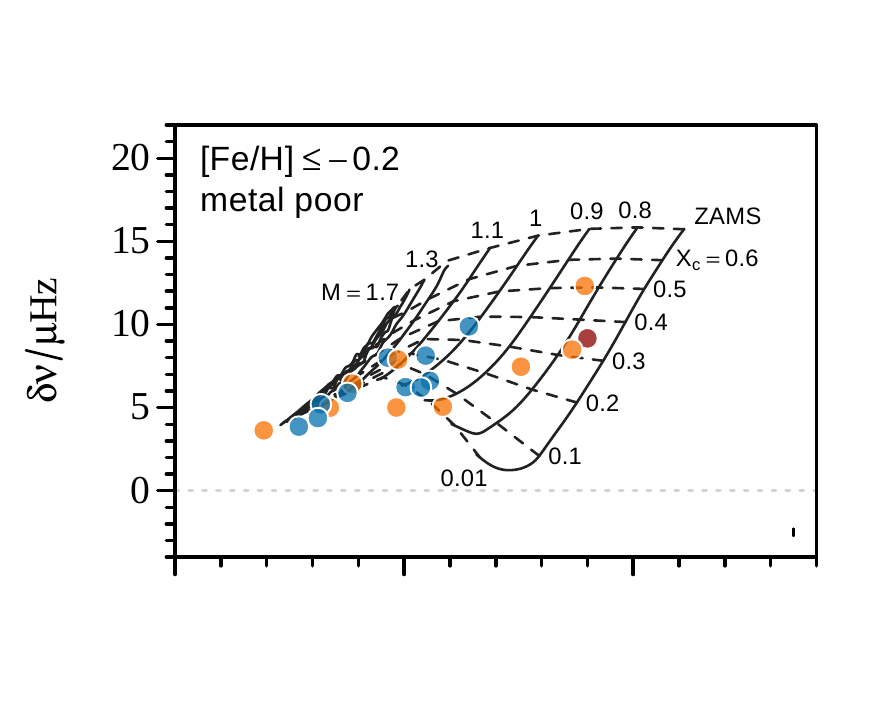}%
        }%
        \adjustbox{trim={\lefttrim} {\bottrim} 0cm 0cm, clip}{%
            \includegraphics[width=0.5\linewidth]{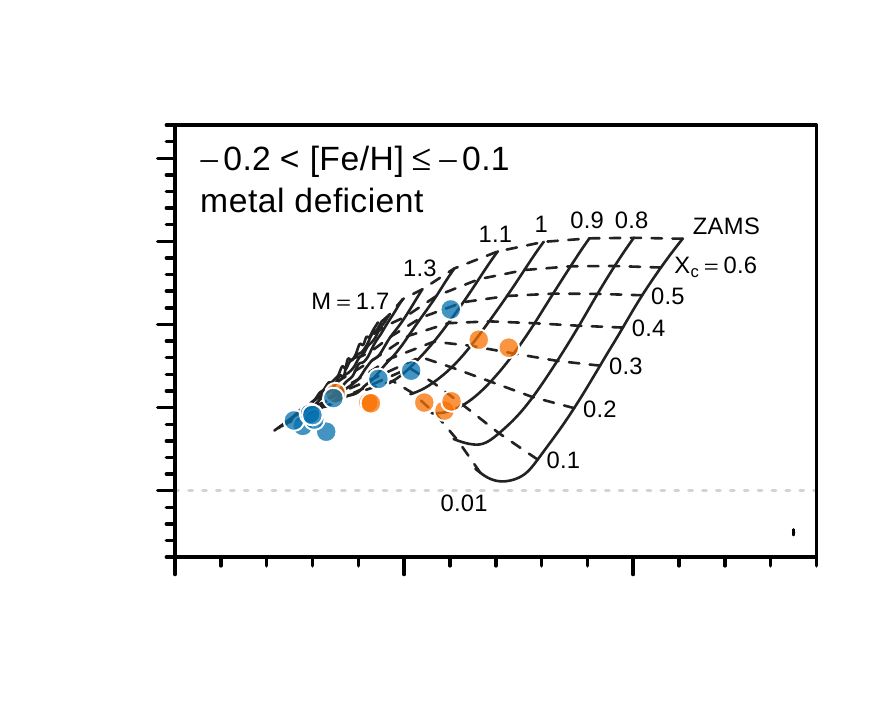}%
        }%
    }%
    \\%
    \makebox[\linewidth][c]{%
        \adjustbox{trim=0cm 0cm 0cm {\righttrim}, clip}{%
            \includegraphics[width=0.5\linewidth]{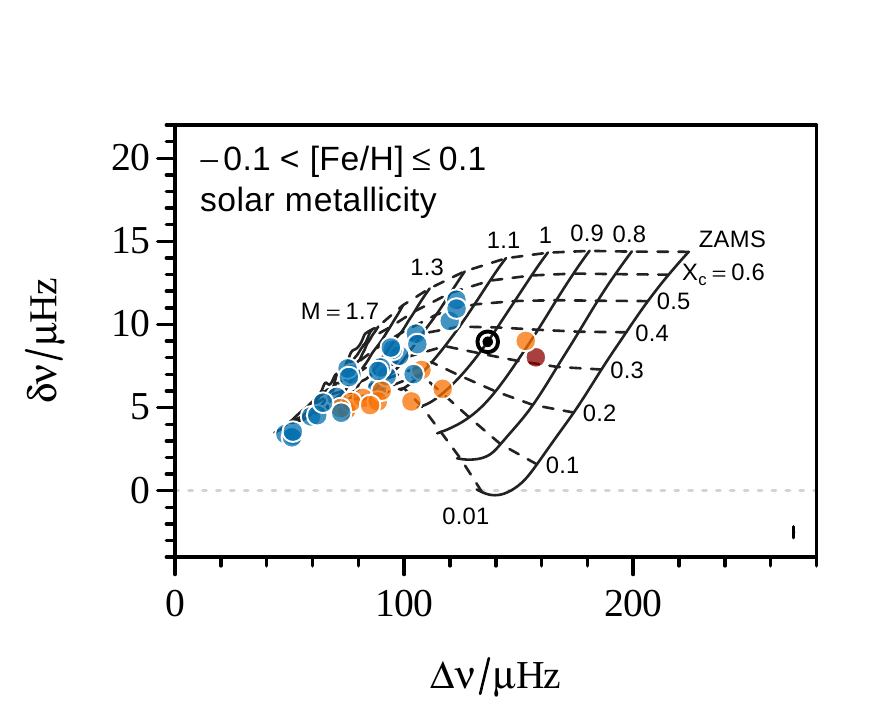}%
        }%
        \adjustbox{trim={\lefttrim} 0cm 0cm {\righttrim}, clip}{%
            \includegraphics[width=0.5\linewidth]{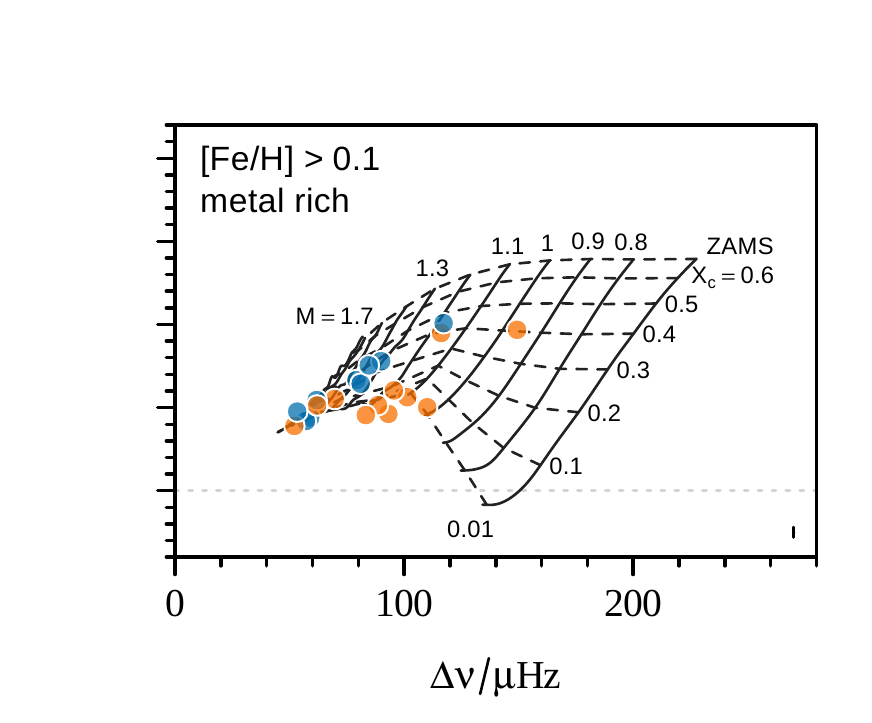}%
        }%
    }%
    \caption{CD diagram showing asteroseismic measurements---the small frequency separation $\delta\nu$ and the the large frequency separation $\Delta\nu$---for the sample of 97 solar-like stars studied here.
    The panels are arranged by metallicity. 
    The points are colored by spectral type: F (blue, ${T_{\text{eff}} > 6000~\text{K}}$), G (orange, ${6000 \geq T_{\text{eff}} > 5200~\text{K}}$), K (red, ${T_{\text{eff}} \leq 5200~\text{K}}$). 
    The cross in the bottom right corner of each panel shows typical uncertainties (see also Figure~\ref{fig:cdf-inputs}). 
    The dotted light gray line indicates where ${\delta\nu = 0}$. 
    The solid lines trace theoretical tracks of stellar evolution, with each track labeled along the top by its mass (in solar units) and connected by the dashed lines that are spaced \mbbb{in} core-hydrogen abundance ($X_c$). 
        \label{fig:CD}}
\end{figure*}

\begin{figure*}%[p]
    \makebox[\linewidth][c]{\adjustbox{trim={0 0 2cm 0},
    clip}{\includegraphics[width=\linewidth, keepaspectratio]{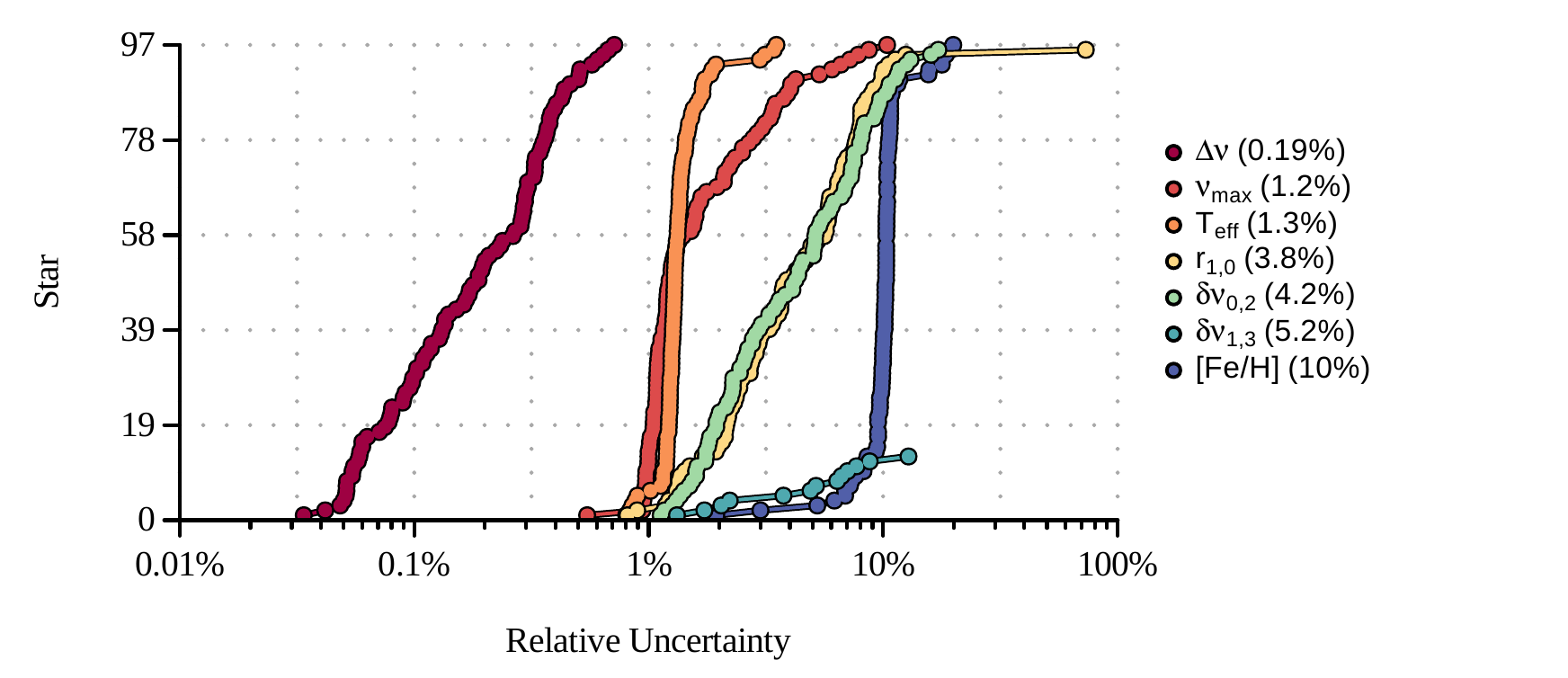}}}
    \caption{Relative uncertainty of each measurement for each of the 97 stars analyzed here. 
    The uncertainties are given in the sense of $\sigma/\mu$ where $\mu$ is the mean of the estimate and $\sigma$ is the standard deviation---except in the case of [Fe/H], where these quantities are instead given for $\exp($[Fe/H]$)$. 
    For each type of measurement, the order of the stars is given in order of increasing relative uncertainty, so for example Star 1 is not necessarily the same star in all the different types of measurements. 
    We note that $\delta\nu_{1,3}$ measurements are only available for 13 of the stars. 
        \label{fig:cdf-inputs}}
%\end{figure}
%
%\begin{figure}
%    \centering
    %\vspace*{3\baselineskip}
    \vspace*{\baselineskip}
    \includegraphics[width=\linewidth, keepaspectratio]{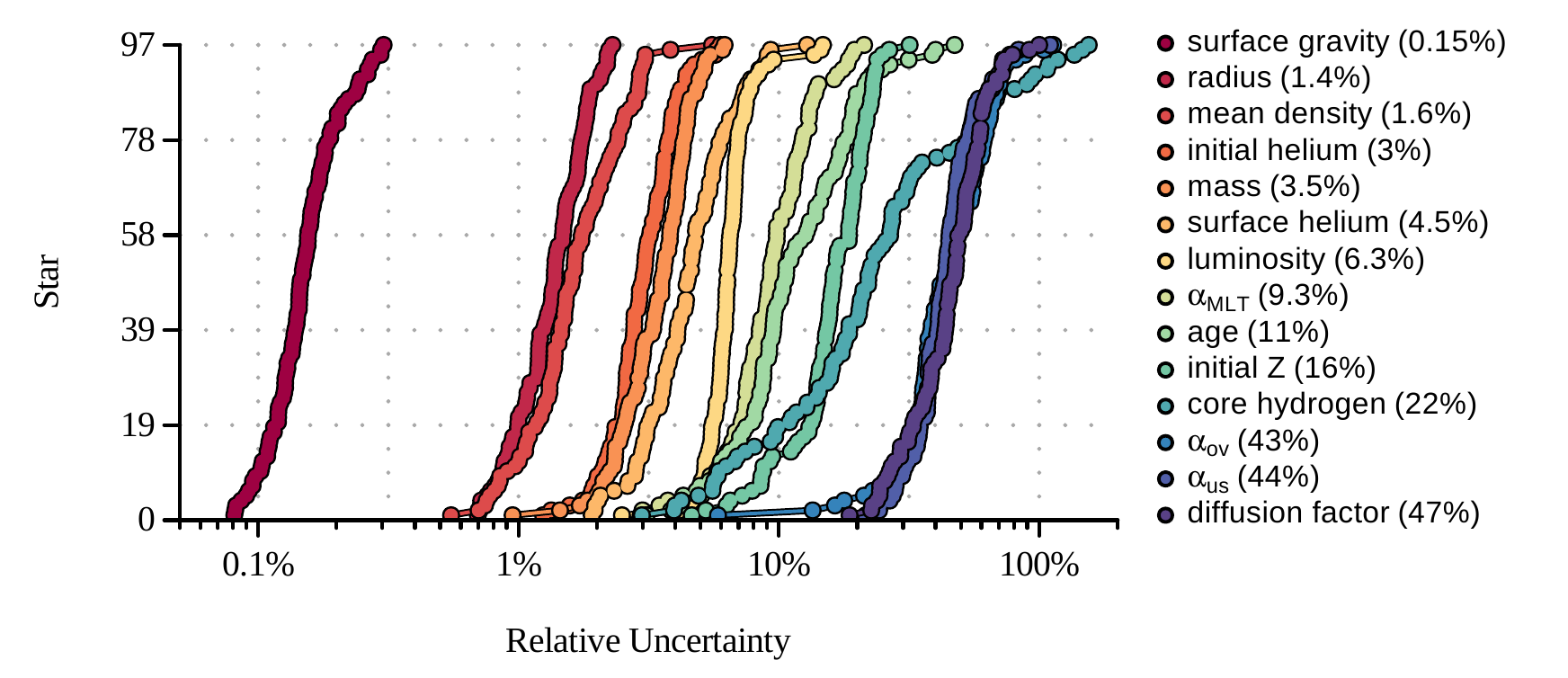}
    \caption{\mb{Cumulative distributions showing the relative uncertainty of each estimated stellar parameter for each of the 97 stars, obtained by running the data in Figure~\ref{fig:cdf-inputs} through SPI. }
        \label{fig:cdf}}
\end{figure*}

We obtained oscillation mode frequencies, \mb{$\nu_{\max}$ values}, effective temperatures, and metallicities for 97 stars observed by the \emph{Kepler} spacecraft \citep{2010Sci...327..977B} from previous studies. 
Data for 31 of the stars came from the \emph{Kepler} Ages project \citep{2015MNRAS.452.2127S, 2016MNRAS.456.2183D}, and 66 were from the \emph{Kepler} LEGACY project \citep{2017ApJ...835..172L, 2017ApJ...835..173S}. 
The spectroscopic parameters of these stars are visualized in Figure~\ref{fig:spectroscopy}. 
We processed the mode frequencies \mb{to obtain asteroseismic frequency separations and ratios} in the same way that we processed the theoretical model frequencies in the previous section. 
These stars are visualized in the so-called CD~diagram \citep{1984srps.conf...11C} in Figure~\ref{fig:CD}. 
The relative uncertainties on the observational constraints for these stars are visualized in Figure~\ref{fig:cdf-inputs}. \mb{This figure furthermore illustrates the observational data we input to SPI. } 
%The relative uncertainties on the observational constraints of these stars are visualized in Figure~\ref{fig:cdf-inputs}; \mb{this figure futhermore shows the data that we input into SPI}. 

For comparison purposes, we additionally obtained luminosity and radius measurements for these stars from the recent \emph{Gaia} Data Release 2 \mb{\citep[DR2,][]{2018A&A...616A...1G, 2018A&A...616A...8A}}. %\citep[DR2,][]{2018A&A...616A...2L}. 
%\emph{Gaia} measurements were not available for 
We \mb{identified these stars using} the 2MASS cross-matched catalog \citep{2018arXiv180809151M} for all except KIC~10514430 and KIC~8379927, which we located using a cone search. 
We note that these measurements have not been corrected for extinction.

\section{Results}

We applied SPI to the sample of 97 stars using the grid of theoretical models that we generated \mb{and the observational constraints shown in Figure~\ref{fig:cdf-inputs}}. %the grid of theoretical models that we generated. 
Random realizations of the observations that fell outside of the range of the grid of models were dropped. 
Figure~\ref{fig:cdf} shows the relative uncertainties in the estimated stellar parameters for this sample. 
The stellar parameters themselves can be found in Table~\ref{tab:parameters} in the Appendix. 
%Our estimates for stars in the LEGACY sample are consistent with those found using other techniques \citep{2017ApJ...835..173S}. 
%We note that except for the Sun, ages of stars are model-dependent. 
Figure~\ref{fig:HR} shows the positions of these stars on the Hertzsprung--Russell diagram using effective temperatures from spectroscopy and predicted luminosities from SPI. 

\subsection{Comparison with \emph{Gaia} data}

\begin{figure}[p]
    \centering
    \makebox[\linewidth][c]{%
        \adjustbox{trim=0cm 1.2cm 0cm 0cm, clip}{%
            \includegraphics[width=0.5\linewidth]{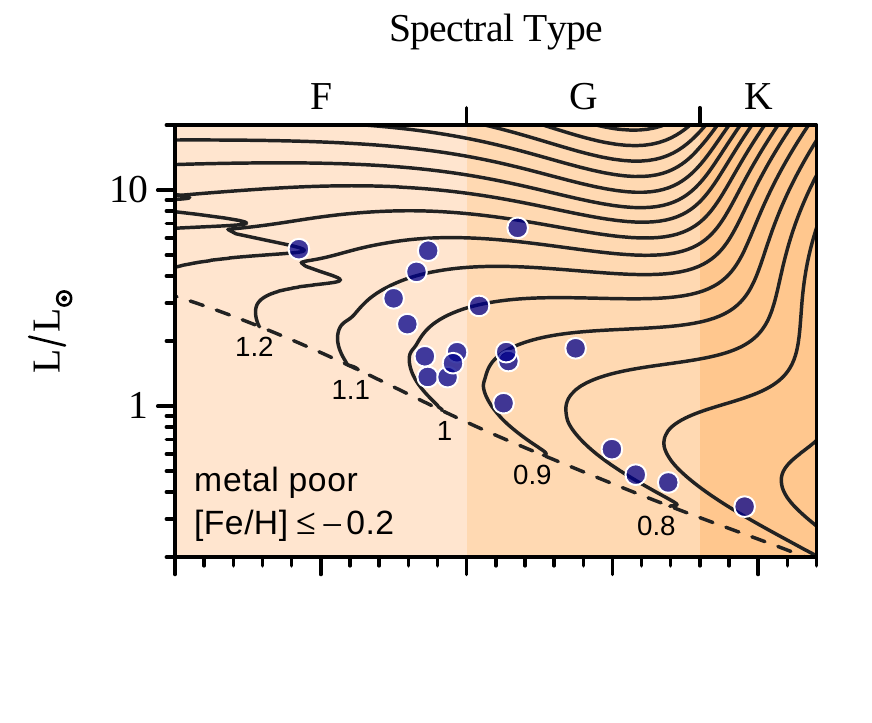}%
        }%
        \adjustbox{trim=1.3cm 1.2cm 0cm 0cm, clip}{%
            \includegraphics[width=0.5\linewidth]{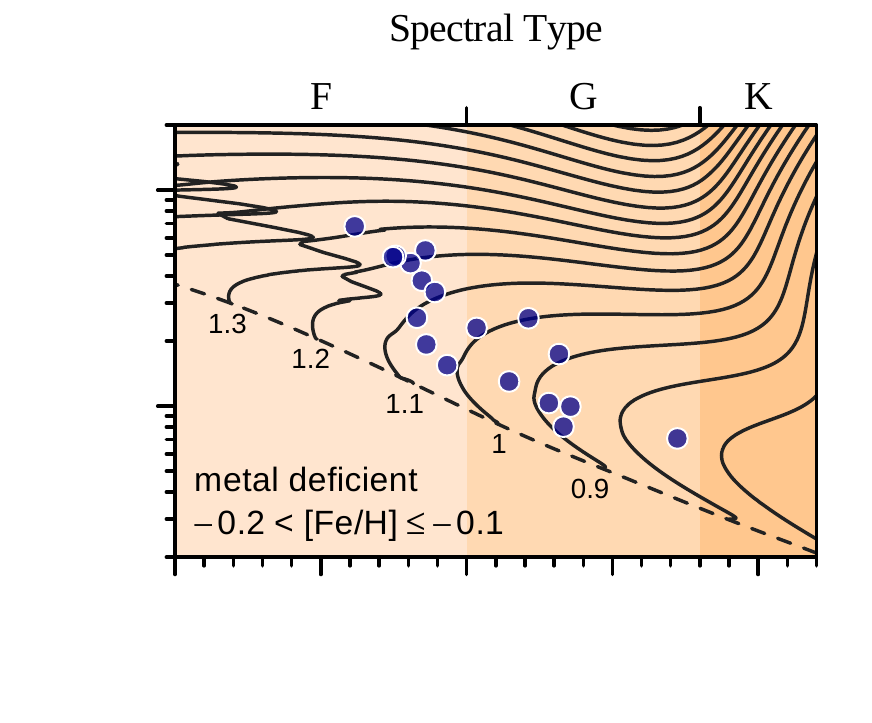}%
        }%
    }%
    \\%
    \makebox[\linewidth][c]{%
        \adjustbox{trim=0cm 0cm 0cm 1cm, clip}{%
            \includegraphics[width=0.5\linewidth]{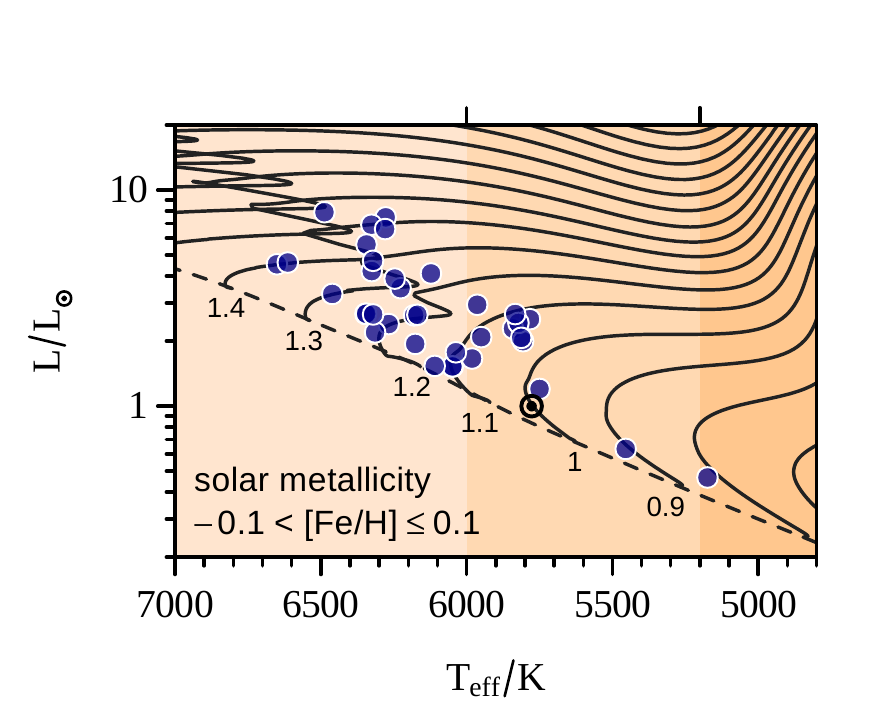}%
        }%
        \adjustbox{trim=1.3cm 0cm 0cm 1cm, clip}{%
            \includegraphics[width=0.5\linewidth]{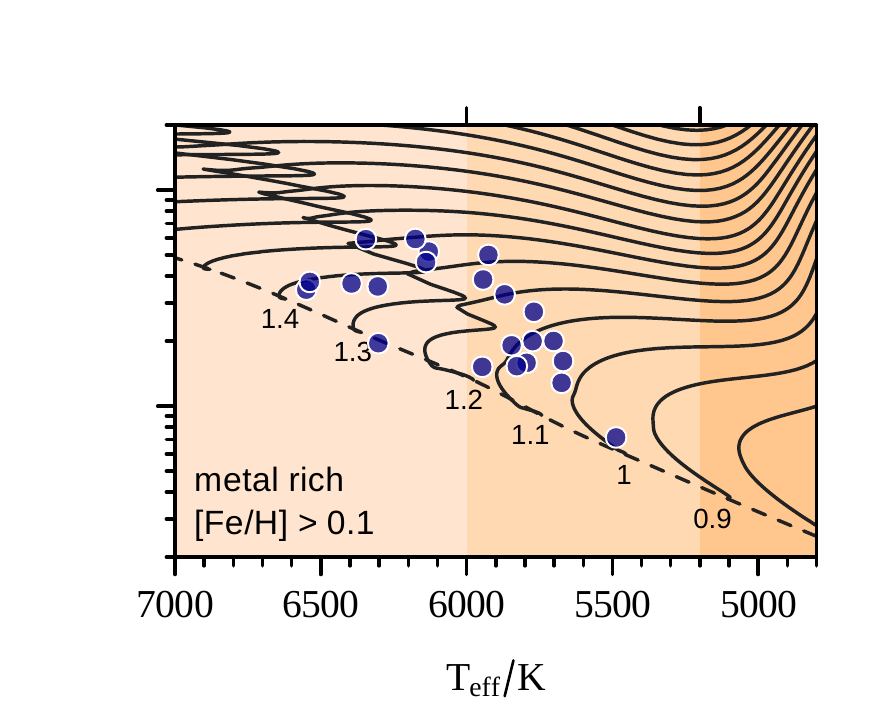}%
        }%
    }%
    \caption{Hertzsprung--Russell diagrams of solar-like stars with temperatures from spectroscopy and predicted luminosities from asteroseismic modeling. 
    The panels are arranged by metallicity, with the lines showing theoretical evolutionary sequences for stars of the corresponding metallicities having the indicated masses (expressed in solar units). 
    The dashed line traces the zero-age main sequence. 
    The background colors indicate the spectral types of the stars (F, G, K). 
    Typical uncertainties are 0.15~$\text{L}_\odot$ and 100~K. 
    %The crosses above the metallicity labels indicate typical uncertainties. 
        \label{fig:HR}}
    
    \vspace*{\baselineskip}
    %SPI vs.\ \emph{Gaia}
    \includegraphics[width=0.33\linewidth]{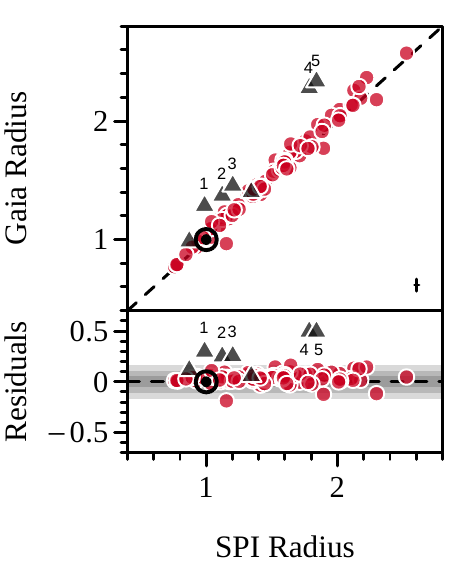}%
    \hspace*{1cm}
    \includegraphics[width=0.33\linewidth]{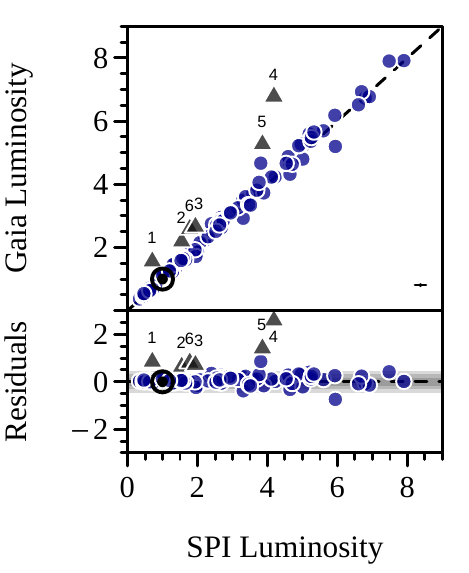}%
    \caption{Comparison of stellar radii (left) and luminosities (right) in solar units from \emph{Gaia} observations against those from stellar modeling using the SPI pipeline. 
    Typical uncertainties are shown as error bars in the lower right corners. 
    Each shade of gray in the residual plots represents an \emph{average\LEt{Please remove the italics here. A and A does not use italics for emphasis.} } $1\sigma$, ranging from one to three. 
    Stars with estimates significantly different at the ${\sigma > 3}$ level are shown as gray triangles. 
    Stars which are outliers in more than one panel are labeled. 
    We note the differences in scale between the left and right panels. 
    The biases in the residuals are consistent with zero ($-0.056\pm0.091$~$\text{R}_\odot$ and $-0.14\pm0.36$~$\text{L}_\odot$)
        \label{fig:gaia}}
\end{figure}

The accuracy of SPI can be gauged by comparing its predictions with observations. 
In Figure~\ref{fig:gaia} we show a comparison of predicted luminosities and radii to those from \emph{Gaia} DR2. Again we find good agreement, despite the systematic errors that may currently be present in the \emph{Gaia} data. 
\citet{2018MNRAS.481L.125S} recently compared their predicted radii to \emph{Gaia} radii for these stars as well and found similar results, although our estimates are in somewhat better agreement, \mb{in the sense that we find fewer discrepant stars}. 
Most likely this is due to the fact that our models incorporate a greater variety of mixing length parameters, which is important for determining stellar radii. 
\mb{We furthermore compare the results from SPI to the BASTA models for these stars \citep{2015MNRAS.452.2127S, 2017ApJ...835..173S} in Figure~\ref{fig:basta}, finding good agreement. } 

\subsection{Testing the seismic scaling relations}

\begin{figure}[p]
    \centering
    \includegraphics[width=0.33\linewidth]{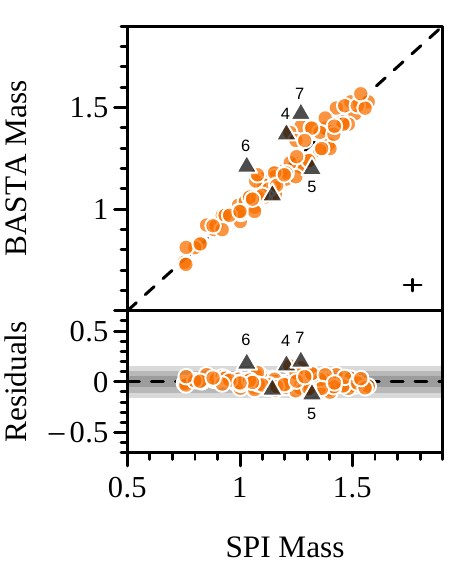}%
    \includegraphics[width=0.33\linewidth]{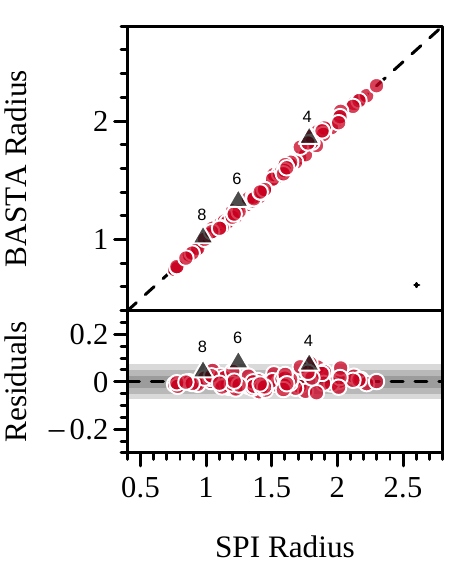}%
    \includegraphics[width=0.33\linewidth]{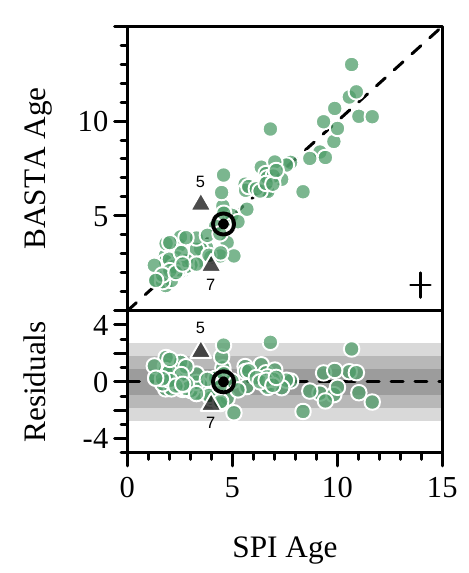}%
    \caption{\mb{Comparison of stellar masses (in solar units), radii (in solar units), and ages (in Gyr) determined via stellar modeling using BASTA against those from stellar modeling using SPI.
    The biases in the residuals are all consistent with zero (${0.001\pm0.061~\text{M}_\odot}$, ${-0.001\pm0.027~\text{R}_\odot}$, and ${-0.1\pm1.0~\text{Gyr}}$).} 
        \label{fig:basta}}
%\end{figure}
%\begin{figure}[p]
    \centering
    \vspace*{3\baselineskip}
    \includegraphics[width=0.33\linewidth]{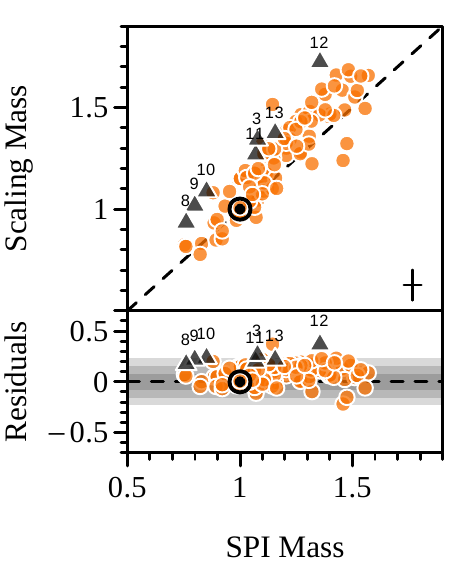}%
    \includegraphics[width=0.33\linewidth]{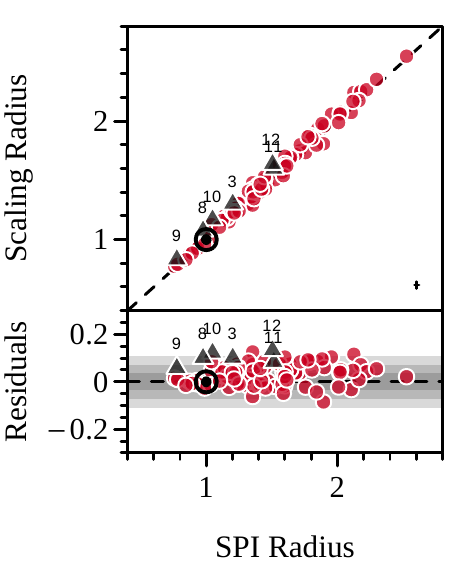}%
    \includegraphics[width=0.33\linewidth]{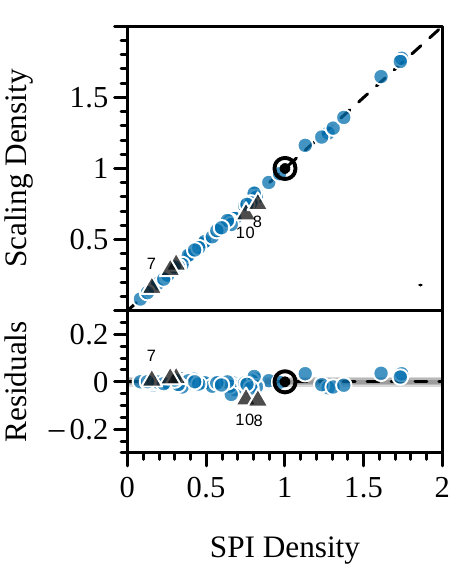}%
    \caption{Comparison of stellar masses, radii, and mean densities (in solar units) determined via the seismic scaling relations (Equations~\ref{eq:scalingR}, \ref{eq:scalingRho}, \ref{eq:scalingM}) against those from stellar modeling using the SPI pipeline.
    The biases in the residuals are all consistent with zero ($0.09\pm0.11$~$\text{M}_\odot$, $0.035\pm0.045$~$\text{R}_\odot$, and $-0.004\pm0.017$~$\rho_\odot$). 
        \label{fig:scaling}}
\end{figure}

The stellar parameters we derive using SPI can be used to check the validity of the seismic scaling relations (Equations~\ref{eq:scalingR}, \ref{eq:scalingRho}, \ref{eq:scalingM}). 
Figure~\ref{fig:scaling} shows a comparison between the masses, radii, and mean densities obtained in these two ways. %obtained via SPI to those from the seismic scaling relations. 
Overall, there is good agreement, albeit with some significant outliers. 
Only one of the labeled outliers, KIC~8478994 is a confirmed planet host (label 9 in the Figure---see Table~\ref{tab:parameters}). 
We note that two of the discrepant stars---KIC~8760414 and KIC~7106245 (labels 8 and 10 in the Figure)---are two of the lowest mass stars and are by far the most metal poor stars of the sample, with [Fe/H] values of approximately ${-0.9}$~dex and ${-1}$~dex, respectively. 
In contrast, the next most metal poor star has a metallicity of approximately ${-0.5}$~dex (\emph{cf}.~Figure~\ref{fig:spectroscopy}). 
However, we find no apparent metallicity dependence in the residuals of any of the scaling relations. 
%\mb{} 

%\subsection{Testing Gaia measurements}

\subsection{Testing systematic errors in spectroscopy}

We performed two sets of tests aimed at gauging the robustness of stellar parameters to unreliable spectroscopic parameters. 
In the first test, we biased all of the spectroscopic measurements by systematically changing their reported metallicities and effective temperatures \mb{(both independently and simultaneously)}. % by 5$\sigma$, i.e., up to 0.5~dex and 500~K, respectively. 
We then measured the extent to which their estimated radii, masses, mean densities, and ages changed with respect to the unperturbed estimates \mb{of the stellar parameter}. 
In the second test, we increased the reported random uncertainties for each star and measured the change to the uncertainties of the resulting stellar parameters. 
%In the third, we biased the metallicities and effective temperatures simultaneously. 

\begin{figure}%[p]
    \centering
    \makebox[\linewidth][c]{%
        \adjustbox{trim=0cm 1.3cm 0cm 0cm, clip}{%
            \includegraphics[width=0.5\linewidth]{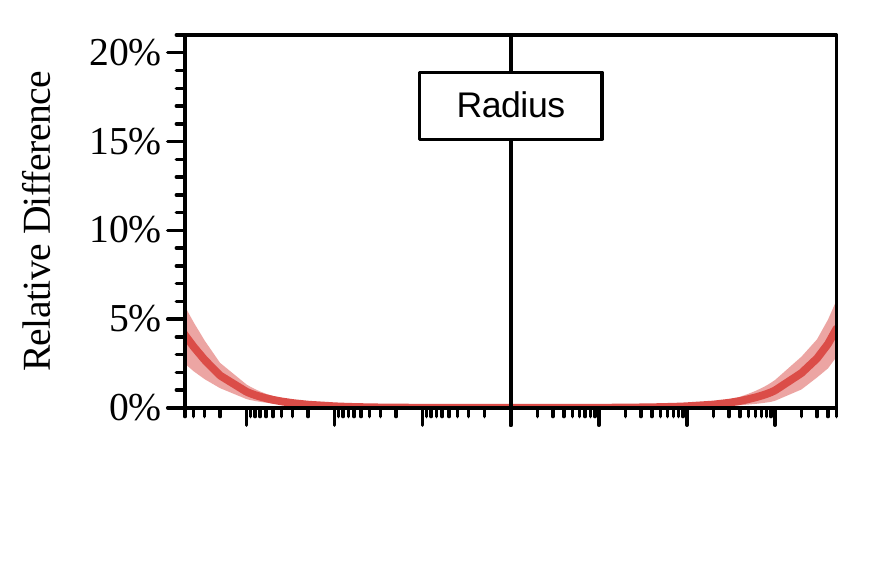}%
        }%
        \adjustbox{trim=1.3cm 1.3cm 0cm 0cm, clip}{%
            \includegraphics[width=0.5\linewidth]{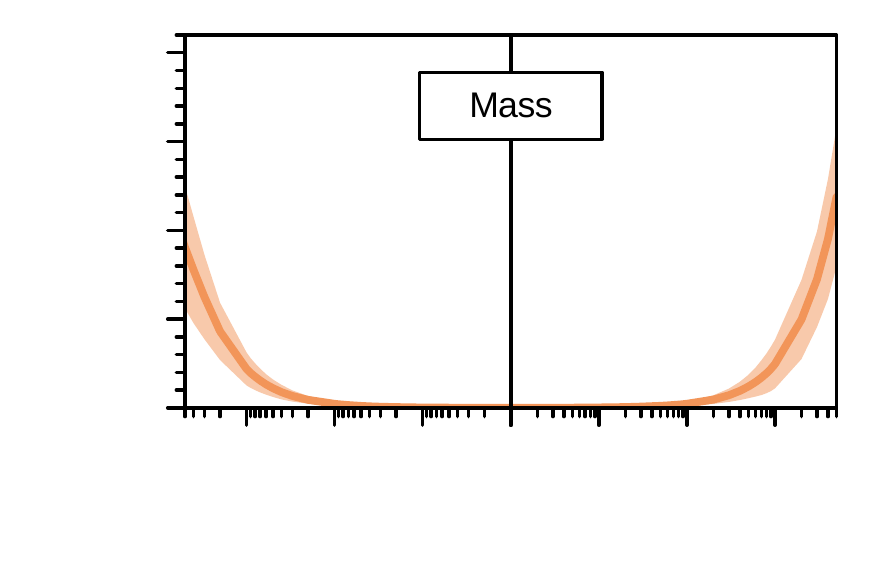}%
        }%
    }%
    \\%
    \makebox[\linewidth][c]{%
        \adjustbox{trim=0cm 0cm 0cm 0cm, clip}{%
            \includegraphics[width=0.5\linewidth]{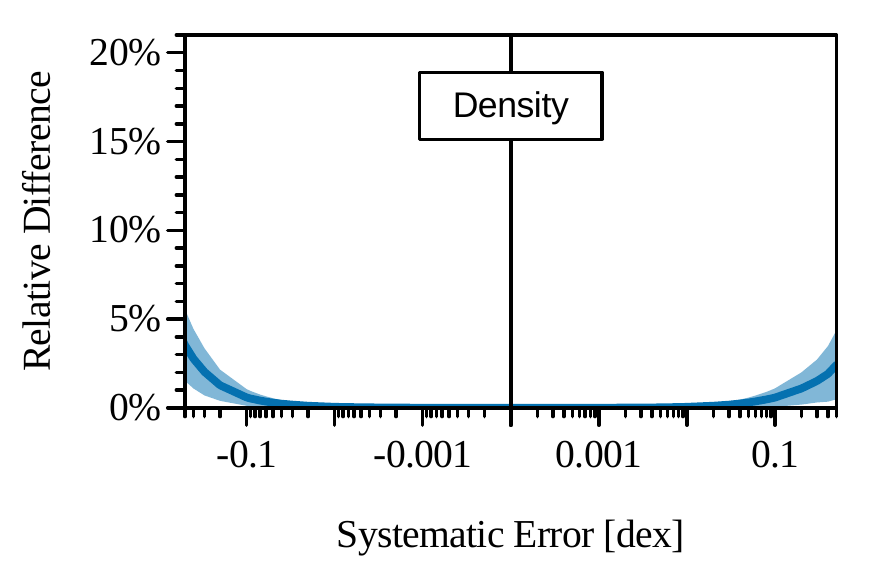}%
        }%
        \adjustbox{trim=1.3cm 0cm 0cm 0cm, clip}{%
            \includegraphics[width=0.5\linewidth]{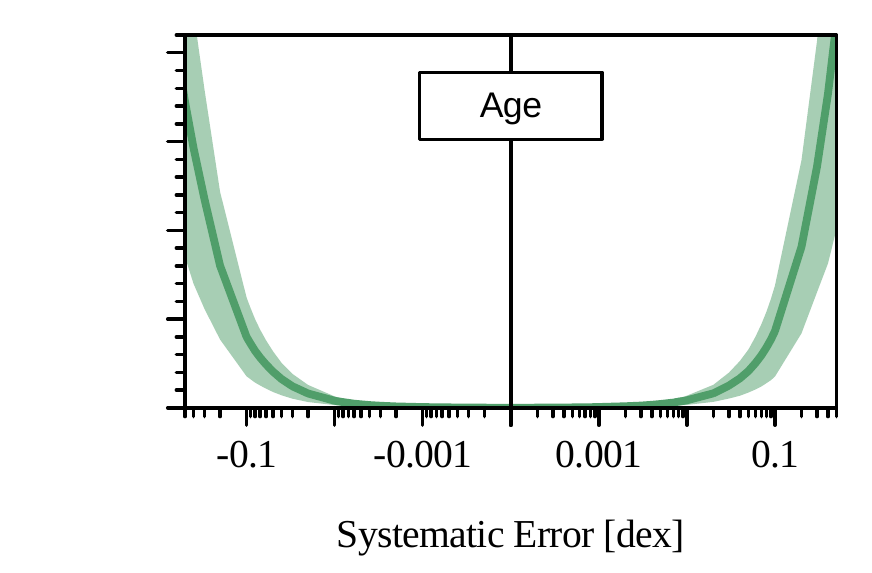}%
        }%
    }%
    \caption{Average relative differences in stellar parameter estimates (radius, mass, density, and stellar age) between estimates made using the reported measurements and estimates made using [Fe/H] values that have been biased with systematic errors ranging from -0.5~dex to 0.5~dex. 
    The lines show the mean values\LEt{Please check I have retained your intended meaning. } and the shaded regions show the standard deviations\LEt{Please check I have retained your intended meaning. } across the 97 stars. 
        \label{fig:bias}} 
%\end{figure}
\vspace*{\baselineskip}
\vfill

%\begin{figure}[p]
    \centering
    \makebox[\linewidth][c]{%
        \adjustbox{trim=0cm 1.3cm 0cm 0cm, clip}{%
            \includegraphics[width=0.5\linewidth]{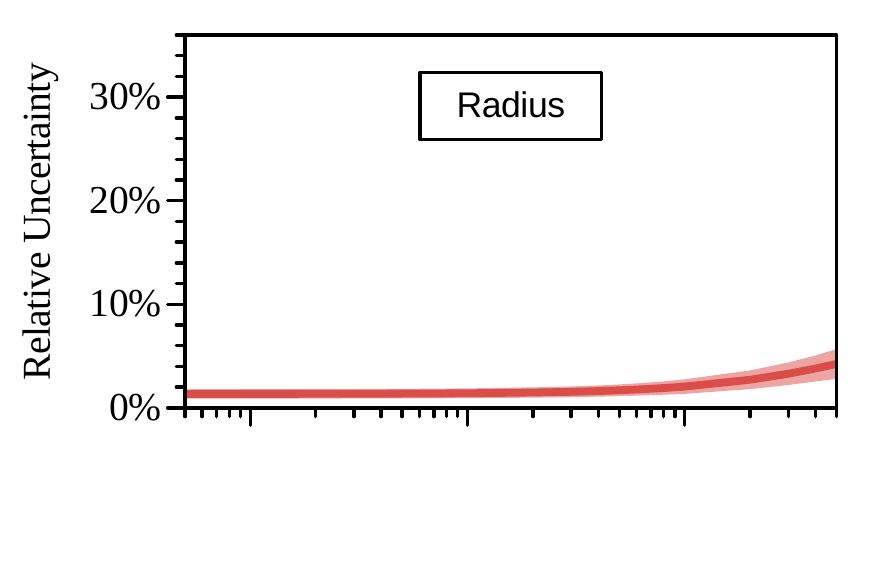}%
        }%
        \adjustbox{trim=1.3cm 1.3cm 0cm 0cm, clip}{%
            \includegraphics[width=0.5\linewidth]{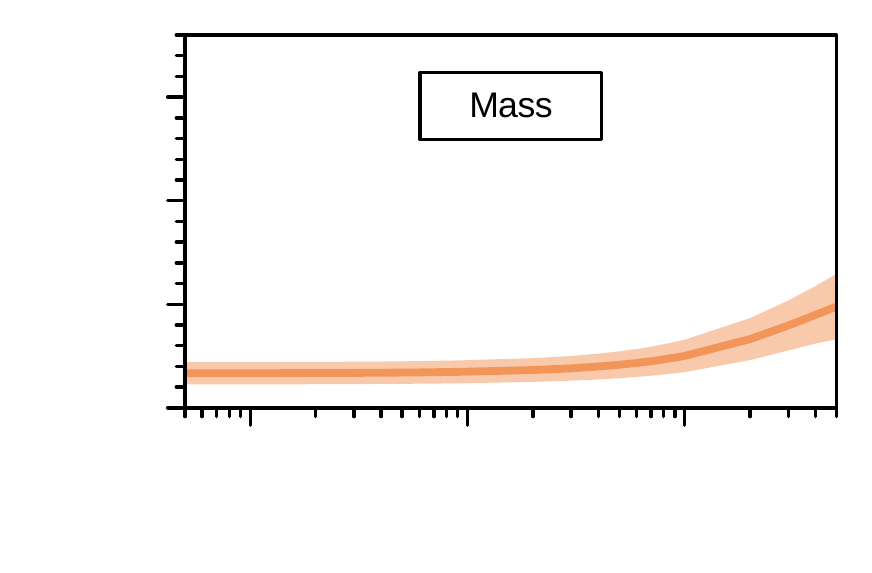}%
        }%
    }%
    \\%
    \makebox[\linewidth][c]{%
        \adjustbox{trim=0cm 0cm 0cm 0cm, clip}{%
            \includegraphics[width=0.5\linewidth]{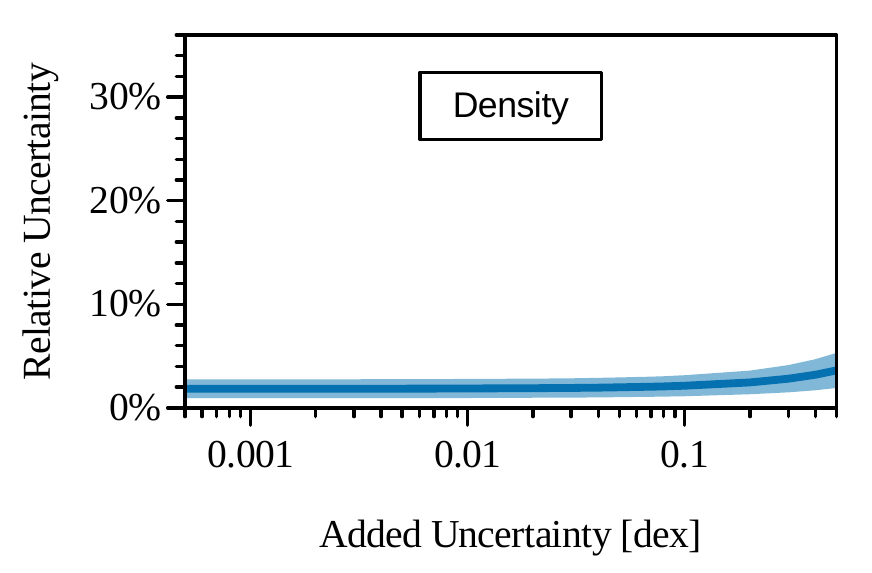}%
        }%
        \adjustbox{trim=1.3cm 0cm 0cm 0cm, clip}{%
            \includegraphics[width=0.5\linewidth]{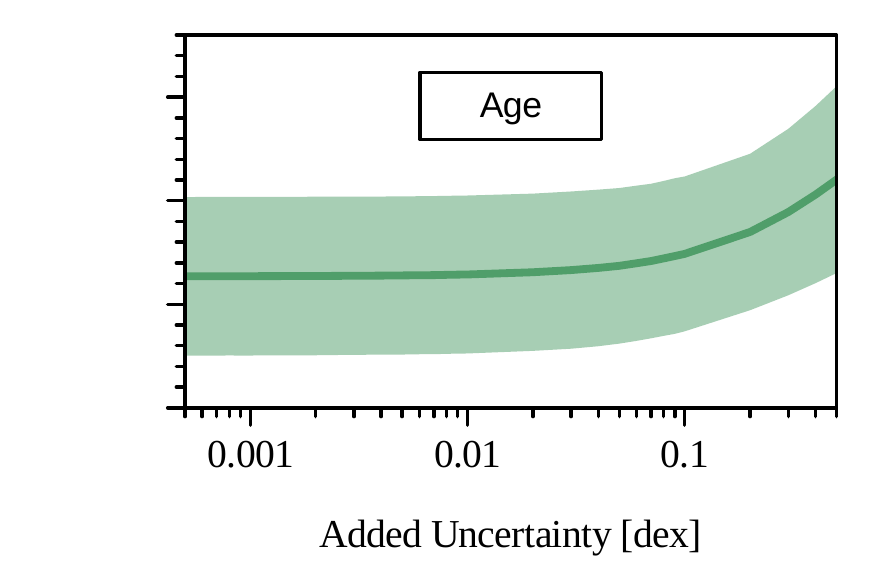}%
        }%
    }%
    \caption{Average relative uncertainties in stellar parameter estimates (radius, mass, mean density, and stellar age) as a function of the amount of additional random uncertainty given to [Fe/H] measurements ranging logarithmically from 0.0005~dex to 0.5~dex.
        \label{fig:imp}}
\end{figure}

\begin{figure}%[p]
    \centering
    \makebox[\linewidth][c]{%
        \adjustbox{trim=0cm 1.3cm 0cm 0cm, clip}{%
            \includegraphics[width=0.5\linewidth]{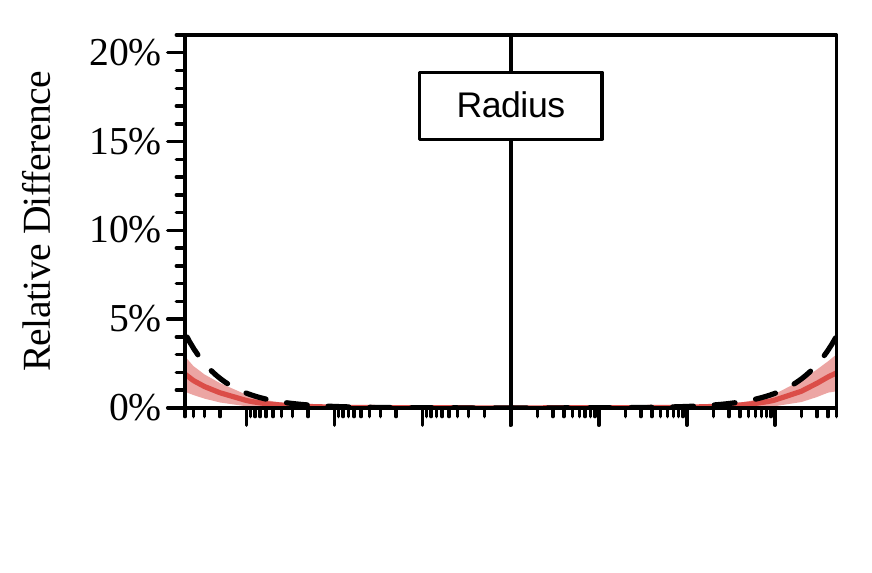}%
        }%
        \adjustbox{trim=1.3cm 1.3cm 0cm 0cm, clip}{%
            \includegraphics[width=0.5\linewidth]{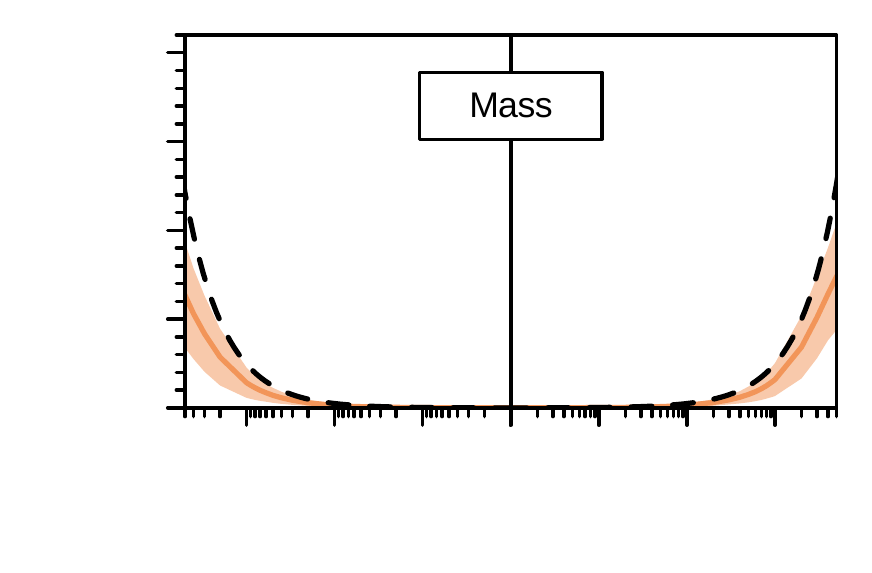}%
        }%
    }%
    \\%
    \makebox[\linewidth][c]{%
        \adjustbox{trim=0cm 0cm 0cm 0cm, clip}{%
            \includegraphics[width=0.5\linewidth]{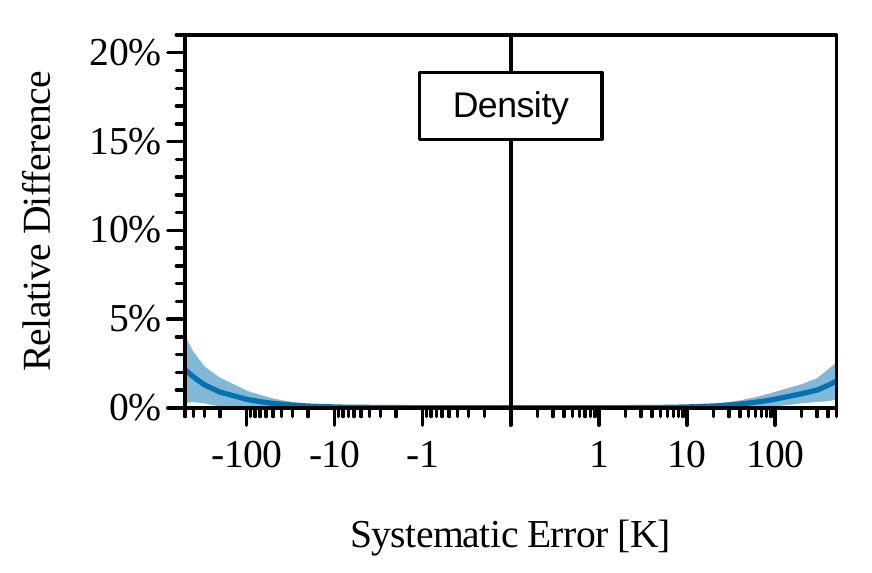}%
        }%
        \adjustbox{trim=1.3cm 0cm 0cm 0cm, clip}{%
            \includegraphics[width=0.5\linewidth]{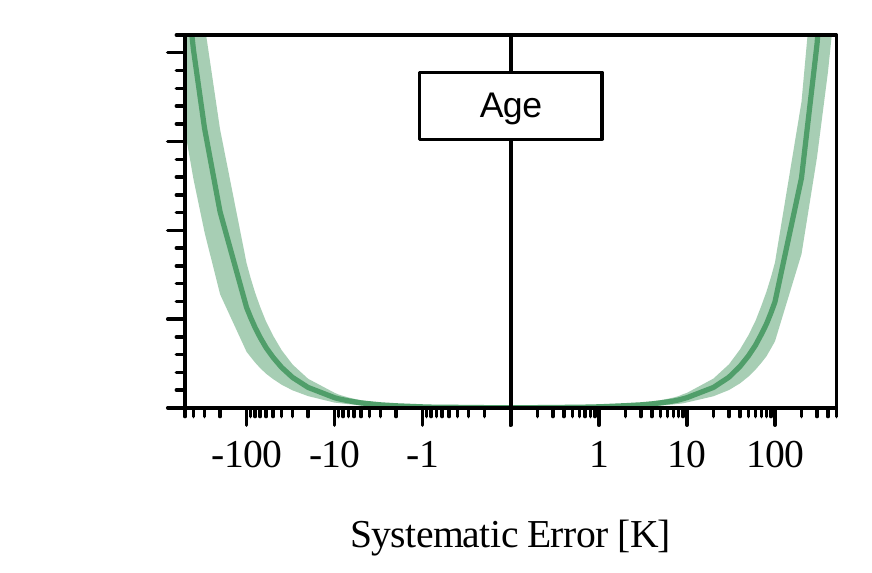}%
        }%
    }%
    \caption{Average relative differences in stellar parameter estimates (radius, mass, mean density, and stellar age) between estimates made using the reported measurements and estimates made using $T_{\text{eff}}$ values that have been biased with systematic errors ranging from -500~K to 500~K. 
    The dashed lines show the relative differences from the seismic scaling relations (Equations~\ref{eq:scalingR}, \ref{eq:scalingRho} and \ref{eq:scalingM}). 
        \label{fig:bias-teff}} 
%\end{figure}
\vspace*{\baselineskip}
\vfill

%\begin{figure}[p]
    \centering
    \makebox[\linewidth][c]{%
        \adjustbox{trim=0cm 1.3cm 0cm 0cm, clip}{%
            \includegraphics[width=0.5\linewidth]{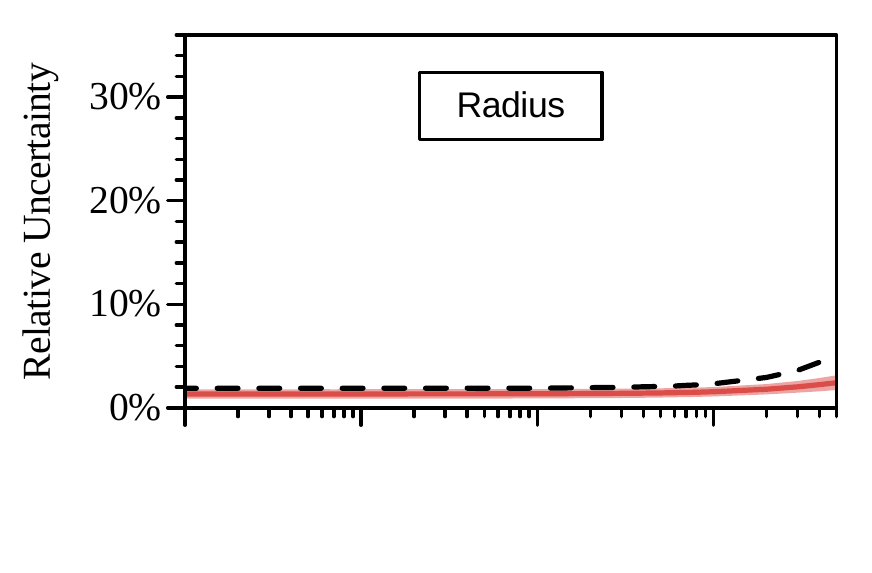}%
        }%
        \adjustbox{trim=1.3cm 1.3cm 0cm 0cm, clip}{%
            \includegraphics[width=0.5\linewidth]{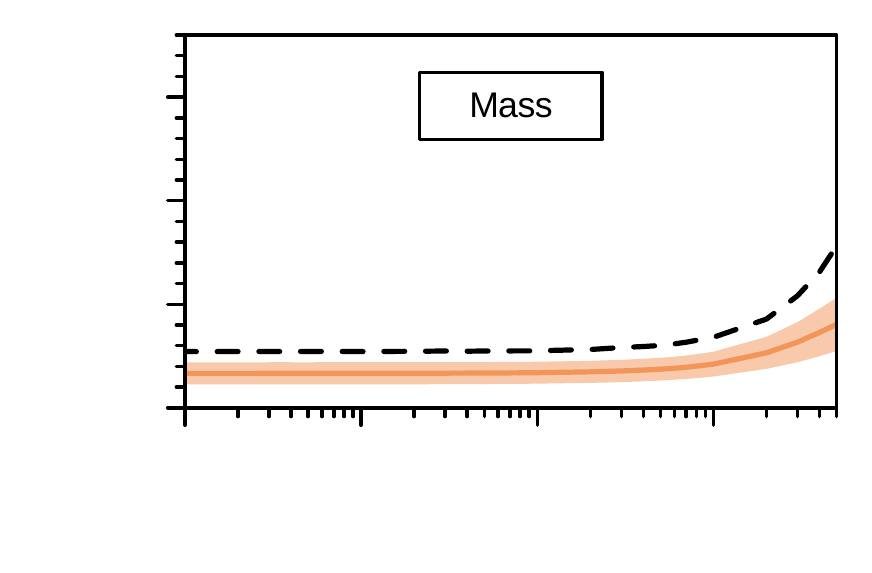}%
        }%
    }%
    \\%
    \makebox[\linewidth][c]{%
        \adjustbox{trim=0cm 0cm 0cm 0cm, clip}{%
            \includegraphics[width=0.5\linewidth]{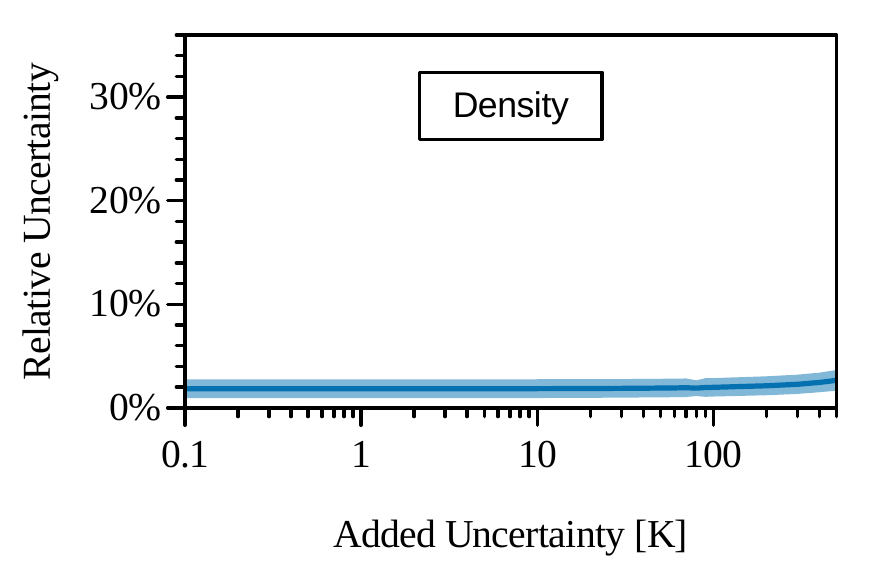}%
        }%
        \adjustbox{trim=1.3cm 0cm 0cm 0cm, clip}{%
            \includegraphics[width=0.5\linewidth]{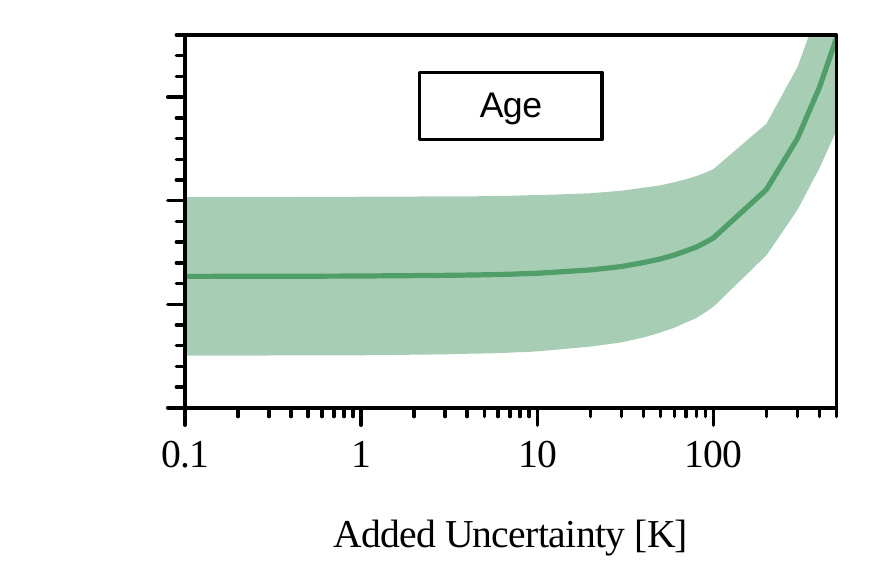}%
        }%
    }%
    \caption{Average relative uncertainties in stellar parameter estimates (radius, mass, mean density, and stellar age) as a function of the amount of additional random uncertainty given to $T_{\text{eff}}$ measurements ranging logarithmically from 0.1~K to 500~K. 
        \label{fig:imp-teff}}
\end{figure}

\begin{figure}[p]
    \centering
    \makebox[\linewidth][c]{%
        \adjustbox{trim=0cm 1.3cm 0cm 0cm, clip}{%
            \includegraphics[width=0.5\linewidth]{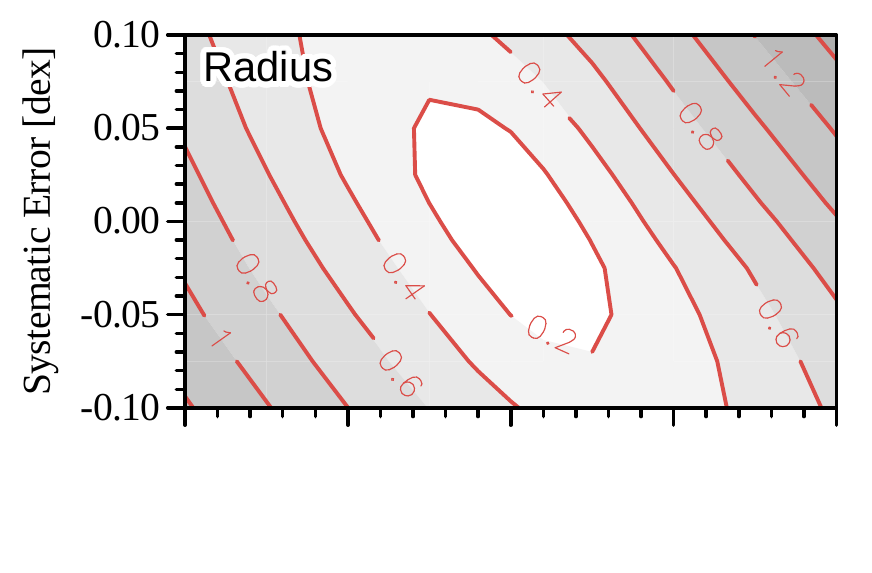}%
        }%
        \adjustbox{trim=1.3cm 1.3cm 0cm 0cm, clip}{%
            \includegraphics[width=0.5\linewidth]{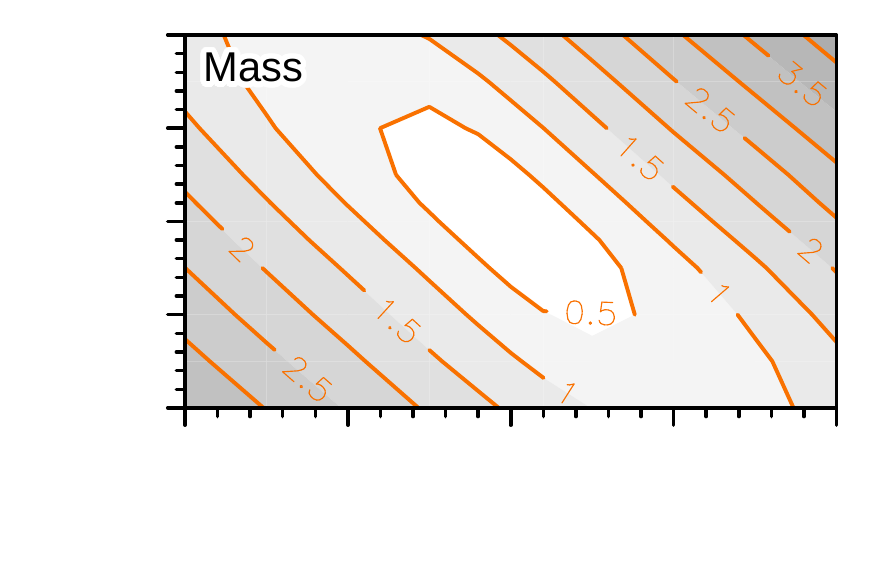}%
        }%
    }%
    \\%
    \makebox[\linewidth][c]{%
        \adjustbox{trim=0cm 0cm 0cm 0cm, clip}{%
            \includegraphics[width=0.5\linewidth]{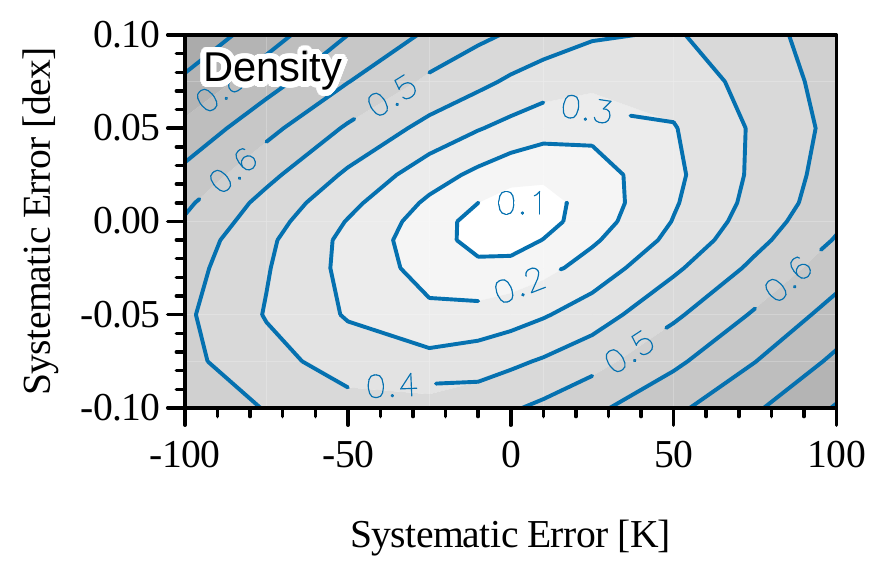}%
        }%
        \adjustbox{trim=1.3cm 0cm 0cm 0cm, clip}{%
            \includegraphics[width=0.5\linewidth]{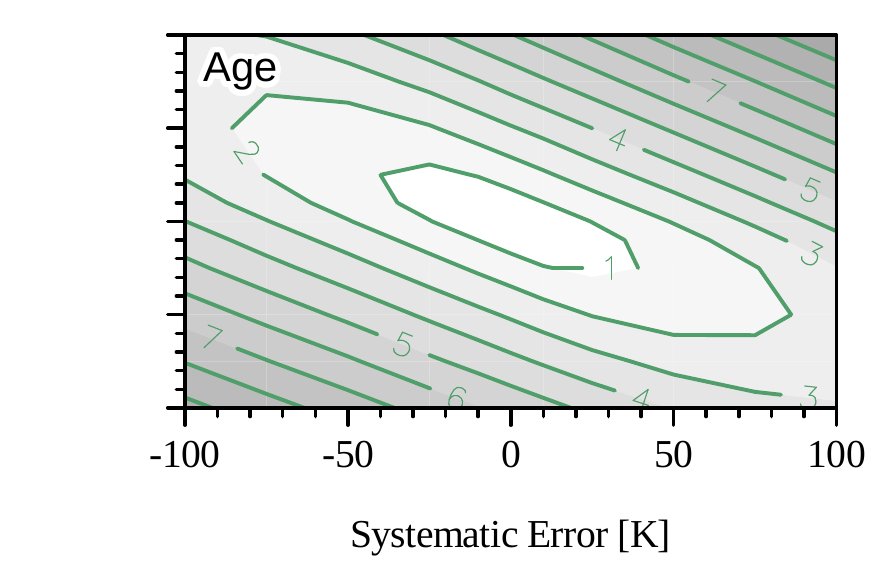}%
        }%
    }%
    \caption{Average relative differences (in percent) in stellar parameter estimates (radius, mass, mean density, and stellar age) between estimates made using the reported measurements and estimates made using \mb{(simultaneously)} systematically biased [Fe/H] and $T_{\text{eff}}$ values.
        \label{fig:bias-both}}

\vspace*{\baselineskip}
\vfill

    \centering
    \makebox[\linewidth][c]{%
        \adjustbox{trim=0cm 1.3cm 0cm 0cm, clip}{%
            \includegraphics[width=0.5\linewidth]{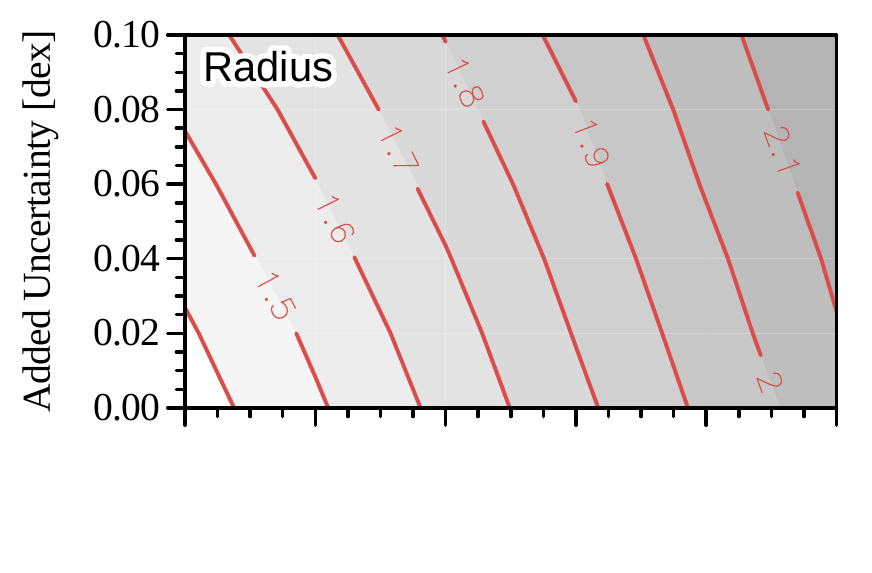}%
        }%
        \adjustbox{trim=1.3cm 1.3cm 0cm 0cm, clip}{%
            \includegraphics[width=0.5\linewidth]{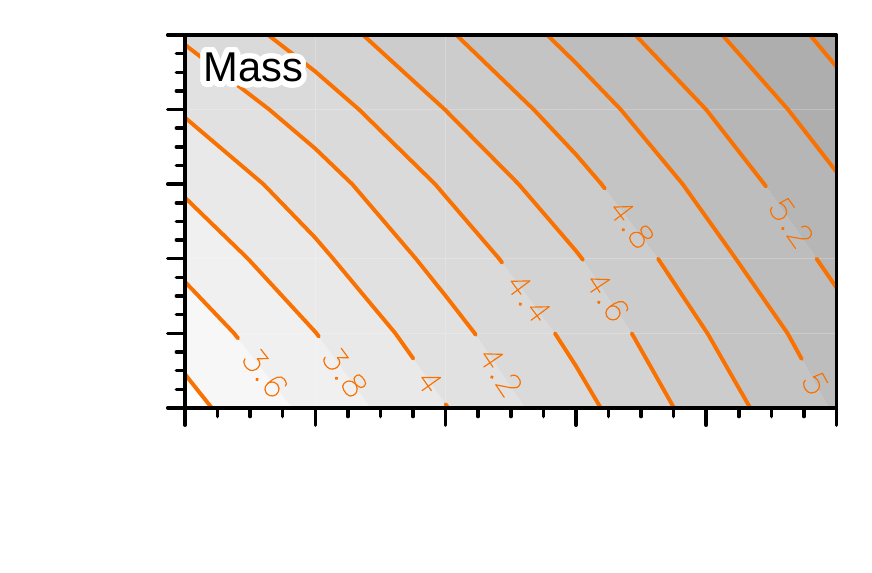}%
        }%
    }%
    \\%
    \makebox[\linewidth][c]{%
        \adjustbox{trim=0cm 0cm 0cm 0cm, clip}{%
            \includegraphics[width=0.5\linewidth]{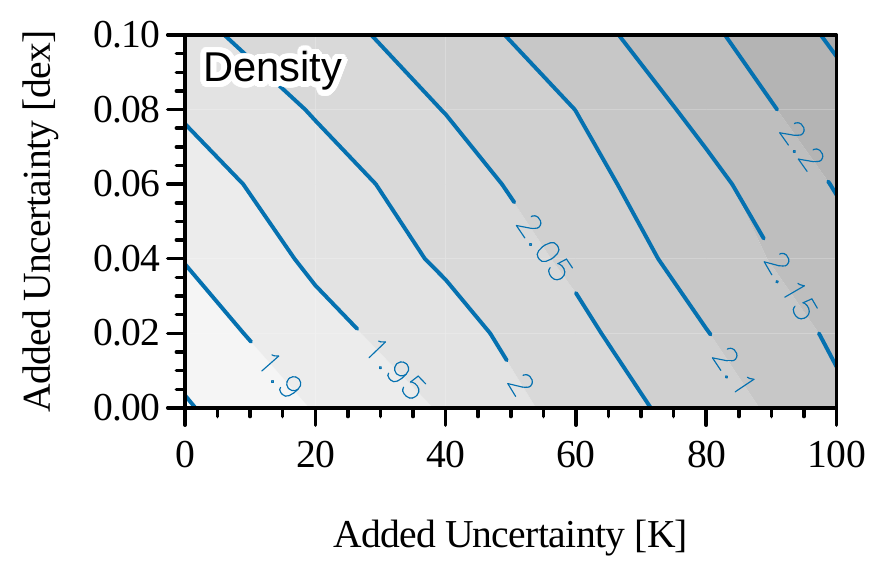}%
        }%
        \adjustbox{trim=1.3cm 0cm 0cm 0cm, clip}{%
            \includegraphics[width=0.5\linewidth]{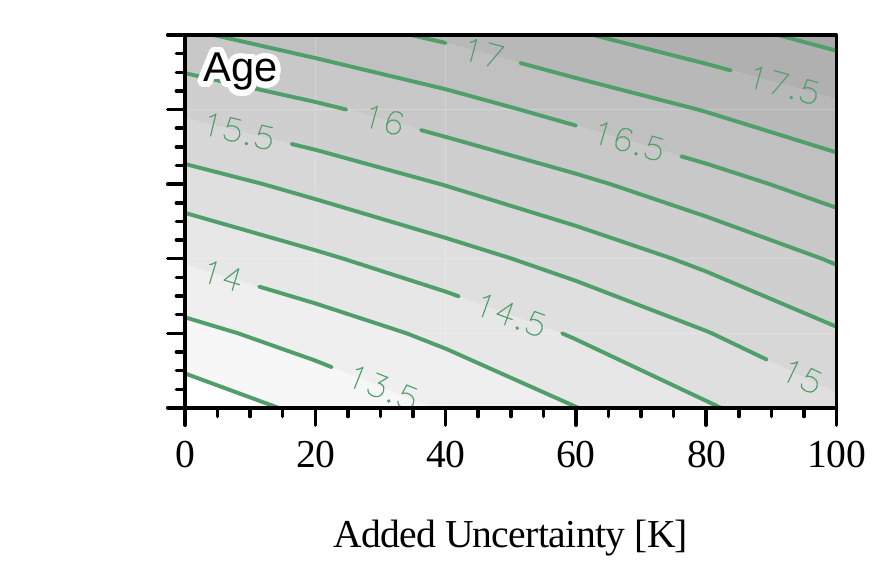}%
        }%
    }%
    \caption{Average relative uncertainties (in percent) in stellar parameter estimates (radius, mass, mean density, and stellar age) as a function of the amount of additional random uncertainty given to [Fe/H] and $T_{\text{eff}}$ measurements.
        \label{fig:imp-both}}
\end{figure}

The results of these experiments are shown in Figures~\ref{fig:bias}, \ref{fig:imp}, \ref{fig:bias-teff}, \ref{fig:imp-teff}, \ref{fig:bias-both}, and \ref{fig:imp-both}. 
There it can be seen that a systematic error of 0.1~dex in [Fe/H] measurements translates on average to differences of only 4\%, 2\%, and 1\% in the resulting stellar ages, masses, and radii, respectively. 
These differences are smaller than the reported relative uncertainties for these quantities ($\sim$~11\%, 3.5\%, 1.4\%). 
Similarly, a systematic error of 100~K in effective temperature results in relative differences of 6\%, 2\%, and 0.4\%.
Furthermore, increasing the reported uncertainty of [Fe/H] measurements by 0.1~dex (100\%) increases the uncertainties of stellar ages, masses, and radii on average by only 0.01~Gyr, 0.02~$\text{M}_\odot$, and 0.01~$\text{R}_\odot$, which are well below the reported uncertainties for these estimates ($\sim$~0.5~Gyr, 0.04~$\text{M}_\odot$, 0.02~$\text{R}_\odot$). 
Similarly, increasing the reported uncertainty of $T_\text{eff}$ by 100~K increases the uncertainties of the estimated stellar parameters by 0.2~Gyr, 0.01~$\text{M}_\odot$, and 0.003~$\text{R}_\odot$. 
Thus, even \mb{relatively large} systematic errors and increases to uncertainty result in essentially the same stellar parameters. 

When biasing [Fe/H] estimates by more than 0.1~dex (1$\sigma$), some of the stellar metallicities went beyond the grid of models and \LEt{Please check I have retained your intended meaning. }were dropped. 
By 0.5~dex (5$\sigma$), only half of the stars remained, and at this point we stopped the experiment. 
We find that in order for the relative differences in the resulting stellar ages, masses, and radii to be equal to their reported uncertainties, [Fe/H] ($T_{\text{eff}}$) measurements would need to be biased by at least 0.24~dex (175~K), 0.14~dex (200~K), and 0.14~dex (300~K), respectively. 

The results regarding the uncertainties are consistent with the theoretical findings of \citet[][their Fig.~7]{2017ApJ...839..116A}. 
Though our analysis was done using frequency ratios to account for the surface term, \citet{BasuKinnane2018} have similarly shown that how the surface term is accounted for does not affect estimates of the global properties (but see also \citealt{2017EPJWC.16003010A}). 

The tests regarding effective temperatures also show these biases propagated through the seismic scaling relations (Equations~\ref{eq:scalingR}, \ref{eq:scalingRho}, \ref{eq:scalingM}). 
We find that estimates obtained using SPI are even more robust to systematic errors than the scaling relations.

\section{Conclusions}
We estimated the stellar parameters of 97 main-sequence and early sub-giant stars observed by the \emph{Kepler} spacecraft using the SPI pipeline. 
We compared the estimates of stellar mass, radius, and mean density that we obtained from stellar modeling for these stars to the estimates from seismic scaling relations and found good agreement. 
We found no evidence in the residuals for a trend with metallicity. 
%Overall, we found good agreement. 
We similarly compared the estimates of stellar radius and luminosity that we obtained from stellar modeling to the measurements from \emph{Gaia} DR2, and again found good agreement. 

We used these stars to test the impact of systematic errors in metallicity and effective temperature on the determination of stellar age, mass, and radius via asteroseismic stellar modeling. 
We found that the resulting stellar parameters from SPI are stable with respect to even \mb{relatively large} systematic errors in the spectroscopic measurements. 
\mb{We emphasize that these results are only valid for asteroseismic modeling performed using SPI; results from other methods may vary. } 
We note that we have not tested the impact on more evolved stars, \mb{such as red giants}, and there too the results may differ. 
\mb{We note also that there may be other sources of systematic errors in the theoretical models, such as the assumed opacities, nuclear reaction rates, and mixture of metals, which we have not examined here. } 
%The masses, radii, and densities determined using SPI are even more resilient to systematic errors in effective temperature measurements than the seismic scaling relations. 
%We note that these results are only valid for asteroseismic modelling using SPI. 
%Thus the results in our paper are robust. 

These results should not be taken to imply that spectroscopic measurements are unimportant. 
As was shown by \citet{2016ApJ...830...31B} and \citet{2017ApJ...839..116A}, metallicity measurements are consistently rated as being highly important for constraining stellar models, as they provide unique information that is not redundant with the oscillation data. 
However, measurements of stellar metallicity are most important in constraining aspects of the models related to the initial chemical composition and mixing processes that take place during the evolution. 
They are, however, comparatively less important in constraining global parameters such as stellar age, mass, and radius \mbb{from existing stellar models}, as these quantities are well-constrained by seismology. 
%As is always the case, the ages and uncertainties we report are relative to the grid of theoretical models. 
%Unknown systematic errors in the models, should they exist, are not propagated into the ages given or their uncertainties. 
%If our understanding of stellar physics were to change, so might the derived ages or their uncertainties. 
%

We conclude that asteroseismic modeling is a very reliable technique for obtaining ages, masses, radii, and other parameters of solar-like oscillators on the main sequence. 
Consequently, the parameters for their exoplanets, being that they are based on these values, are also robust to such systematics. 
%\newpage

%

\onecolumn

%\afterpage{\clearpage}

\appendix 
\onecolumn
\section{Stellar Parameters}
\renewcommand*\footnoterule{}
%\begin{longtable}[]
%    \centering%\scriptsize
%    \begin{tabular}{c|ccccc}
%\begin{center}
\begin{longtable}[c]{c|cccccc}
\caption{Stellar parameters for the \emph{Kepler} Ages and LEGACY stars from the SPI pipeline: age, mass, radius, mean density, luminosity, and number of confirmed exoplanets. The labels next to some of the KIC numbers refer to the outliers highlighted in Figure~\ref{fig:scaling} and \ref{fig:gaia}. The footnotes in the planets column refer to the publication that announced or confirmed their discovery. \label{tab:parameters}}\\
\multicolumn{1}{c|}{KIC} & 
\multicolumn{1}{c}{$\tau/\text{Gyr}$} & 
\multicolumn{1}{c}{$M/\text{M}_\odot$} & 
\multicolumn{1}{c}{$R/\text{R}_\odot$} & 
\multicolumn{1}{c}{$\rho/\rho_\odot$} & 
\multicolumn{1}{c}{$L/\text{L}_\odot$} & 
\multicolumn{1}{c}{Planets} \\\hline\hline 
\endfirsthead
\multicolumn{6}{c}%
{{\bfseries \tablename\ \thetable{}} (continued from previous page) \vspace*{3mm}} \\
\multicolumn{1}{c|}{KIC} & 
\multicolumn{1}{c}{$\tau/\text{Gyr}$} & 
\multicolumn{1}{c}{$M/\text{M}_\odot$} & 
\multicolumn{1}{c}{$R/\text{R}_\odot$} & 
\multicolumn{1}{c}{$\rho/\rho_\odot$} & 
\multicolumn{1}{c}{$L/\text{L}_\odot$} & 
\multicolumn{1}{c}{Planets} \\\hline \hline 
\endhead
%\hline\multicolumn{6}{r}{{ Continued on next page}} \\ \hline
\hline%\hline
\endfoot
\hline%\hline
%\hline\multicolumn{6}{l}{{ }}\\ \hline
\endlastfoot
%KIC & $\tau/\text{Gyr}$ & $M/\text{M}_\odot$ & $R/\text{R}_\odot$ & $L/\text{L}_\odot$ & Label \\\hline\hline
1435467   	&	$2.03\pm0.37$   	&	$1.396\pm0.033$   	&	$1.712\pm0.018$   	&	$0.2789\pm0.0045$   	&	$4.22\pm0.21$   	&	 \Tstrut{}\\
2837475   	&	$1.34\pm0.51$   	&	$1.455\pm0.061$   	&	$1.638\pm0.022$   	&	$0.331\pm0.010$   	&	$4.62\pm0.30$   	&	 \\
3425851   	&	$3.77\pm0.58$   	&	$1.145\pm0.049$   	&	$1.354\pm0.020$   	&	$0.4625\pm0.0062$   	&	$2.67\pm0.17$   	&	 \\
3427720   	&	$2.34\pm0.23$   	&	$1.108\pm0.032$   	&	$1.121\pm0.010$   	&	$0.7904\pm0.0079$   	&	$1.527\pm0.079$   	&	 \\
3456181   	&	$2.62\pm0.37$   	&	$1.428\pm0.062$   	&	$2.125\pm0.035$   	&	$0.1489\pm0.0027$   	&	$6.79\pm0.38$   	&	 \\
3544595   	&	$6.63\pm0.64$   	&	$0.885\pm0.038$   	&	$0.924\pm0.016$   	&	$1.128\pm0.025$   	&	$0.804\pm0.058$   	&	2\footnotemark[1]\footnotetext[1]{\citealt{2014ApJS..210...20M}} \\
3632418   	&	$3.40\pm0.37$   	&	$1.271\pm0.047$   	&	$1.849\pm0.026$   	&	$0.2008\pm0.0020$   	&	$4.59\pm0.33$   	&	1\footnotemark[2]\footnotetext[2]{\citealt{2012ApJ...746..123H}} \\
3656476   	&	$9.16\pm0.80$   	&	$1.079\pm0.027$   	&	$1.318\pm0.017$   	&	$0.4749\pm0.0095$   	&	$1.62\pm0.11$   	&	 \\
3735871   	&	$1.81\pm0.26$   	&	$1.108\pm0.032$   	&	$1.109\pm0.010$   	&	$0.816\pm0.014$   	&	$1.535\pm0.063$   	&	 \\
4141376   	&	$3.48\pm0.56$   	&	$0.993\pm0.041$   	&	$1.038\pm0.013$   	&	$0.897\pm0.031$   	&	$1.365\pm0.094$   	&	1\footnotemark[3]\footnotetext[3]{\citealt{2018AJ....155..203H}} \\
4143755   	&	$10.56\pm0.52$   	&	$0.898\pm0.019$   	&	$1.414\pm0.013$   	&	$0.3211\pm0.0036$   	&	$1.85\pm0.11$   	&	 \\
4349452   	&	$2.47\pm0.55$   	&	$1.206\pm0.055$   	&	$1.312\pm0.023$   	&	$0.535\pm0.011$   	&	$2.40\pm0.15$   	&	3\footnotemark[1]$^,$\footnotemark[4]\footnotetext[4]{\citealt{2012MNRAS.421.2342S}} \\
4914423   	&	$5.60\pm0.68$   	&	$1.117\pm0.059$   	&	$1.468\pm0.021$   	&	$0.3546\pm0.0058$   	&	$2.29\pm0.14$   	&	2\footnotemark[1] \\
4914923   	&	$6.37\pm0.33$   	&	$1.123\pm0.033$   	&	$1.388\pm0.014$   	&	$0.4207\pm0.0042$   	&	$2.00\pm0.11$   	&	 \\
5094751   	&	$5.63\pm0.79$   	&	$1.072\pm0.041$   	&	$1.353\pm0.019$   	&	$0.4346\pm0.0098$   	&	$2.09\pm0.13$   	&	2\footnotemark[1] \\
5184732   	&	$4.84\pm0.35$   	&	$1.205\pm0.022$   	&	$1.3427\pm0.0085$   	&	$0.5034\pm0.0050$   	&	$1.912\pm0.098$   	&	 \\
5773345   	&	$2.24\pm0.50$   	&	$1.510\pm0.044$   	&	$2.017\pm0.021$   	&	$0.1841\pm0.0026$   	&	$5.19\pm0.33$   	&	 \\
5866724   	&	$2.53\pm0.42$   	&	$1.273\pm0.032$   	&	$1.423\pm0.017$   	&	$0.4434\pm0.0064$   	&	$2.648\pm0.098$   	&	3\footnotemark[5]\footnotetext[5]{\citealt{2013ApJ...766..101C}} \\
5950854   	&	$9.8\pm1.1$   	&	$0.984\pm0.044$   	&	$1.236\pm0.023$   	&	$0.524\pm0.012$   	&	$1.62\pm0.14$   	&	 \\
6106415   	&	$4.98\pm0.50$   	&	$1.069\pm0.038$   	&	$1.219\pm0.018$   	&	$0.596\pm0.011$   	&	$1.78\pm0.12$   	&	 \\
6116048   	&	$6.80\pm0.63$   	&	$1.002\pm0.031$   	&	$1.218\pm0.014$   	&	$0.5602\pm0.0080$   	&	$1.78\pm0.11$   	&	 \\
6196457   	&	$4.5\pm1.0$   	&	$1.268\pm0.069$   	&	$1.756\pm0.029$   	&	$0.2359\pm0.0042$   	&	$3.29\pm0.25$   	&	3\footnotemark[6]$^,$\footnotetext[6]{\citealt{2014ApJS..210...25X}}\footnotemark[7]\footnotetext[7]{\citealt{2015ApJ...808..126V}} \\
6225718$^{\text{\hyperref[fig:scaling]{13}}}$   	&	$3.02\pm0.51$   	&	$1.156\pm0.043$   	&	$1.234\pm0.016$   	&	$0.6230\pm0.0082$   	&	$2.19\pm0.13$   	&	 \\
6278762   	&	$10.90\pm0.86$   	&	$0.753\pm0.012$   	&	$0.7588\pm0.0045$   	&	$1.741\pm0.023$   	&	$0.343\pm0.022$   	&	5\footnotemark[8]\footnotetext[8]{\citealt{2015ApJ...799..170C}} \\
6508366   	&	$2.49\pm0.19$   	&	$1.492\pm0.052$   	&	$2.177\pm0.032$   	&	$0.1446\pm0.0024$   	&	$6.91\pm0.47$   	&	 \\
6521045   	&	$6.44\pm0.35$   	&	$1.091\pm0.031$   	&	$1.525\pm0.011$   	&	$0.3090\pm0.0034$   	&	$2.43\pm0.14$   	&	3\footnotemark[1] \\
6603624   	&	$7.77\pm0.45$   	&	$1.065\pm0.023$   	&	$1.173\pm0.012$   	&	$0.668\pm0.016$   	&	$1.283\pm0.059$   	&	 \\
6679371   	&	$2.13\pm0.12$   	&	$1.570\pm0.030$   	&	$2.222\pm0.017$   	&	$0.1432\pm0.0018$   	&	$7.90\pm0.44$   	&	 \\
6933899   	&	$6.61\pm0.28$   	&	$1.104\pm0.023$   	&	$1.599\pm0.012$   	&	$0.2712\pm0.0020$   	&	$2.67\pm0.18$   	&	 \\
7103006   	&	$2.09\pm0.45$   	&	$1.482\pm0.056$   	&	$1.954\pm0.026$   	&	$0.1985\pm0.0038$   	&	$5.60\pm0.38$   	&	 \\
7106245$^{\text{\hyperref[fig:scaling]{10}}}$   	&	$8.36\pm0.72$   	&	$0.851\pm0.025$   	&	$1.048\pm0.014$   	&	$0.752\pm0.017$   	&	$1.363\pm0.098$   	&	 \\
7199397   	&	$3.63\pm0.33$   	&	$1.308\pm0.034$   	&	$2.526\pm0.028$   	&	$0.0821\pm0.0014$   	&	$6.69\pm0.29$   	&	 \\
7206837   	&	$1.80\pm0.55$   	&	$1.400\pm0.041$   	&	$1.584\pm0.016$   	&	$0.3516\pm0.0052$   	&	$3.58\pm0.18$   	&	 \\
7296438   	&	$6.60\pm0.40$   	&	$1.158\pm0.034$   	&	$1.405\pm0.017$   	&	$0.4187\pm0.0045$   	&	$2.00\pm0.11$   	&	 \\
7510397$^{\text{\hyperref[fig:scaling]{4}}}$   	&	$3.97\pm0.35$   	&	$1.206\pm0.030$   	&	$1.784\pm0.015$   	&	$0.2127\pm0.0017$   	&	$4.18\pm0.28$   	&	 \\
7670943   	&	$2.36\pm0.47$   	&	$1.309\pm0.046$   	&	$1.447\pm0.019$   	&	$0.4336\pm0.0055$   	&	$3.31\pm0.27$   	&	 \\
7680114   	&	$7.57\pm0.85$   	&	$1.108\pm0.033$   	&	$1.411\pm0.022$   	&	$0.398\pm0.011$   	&	$2.07\pm0.14$   	&	 \\
7771282   	&	$3.28\pm0.56$   	&	$1.353\pm0.064$   	&	$1.681\pm0.032$   	&	$0.2866\pm0.0049$   	&	$3.89\pm0.23$   	&	 \\
7871531   	&	$9.34\pm0.58$   	&	$0.830\pm0.022$   	&	$0.870\pm0.011$   	&	$1.274\pm0.025$   	&	$0.632\pm0.040$   	&	 \\
7940546   	&	$2.80\pm0.20$   	&	$1.317\pm0.026$   	&	$1.899\pm0.018$   	&	$0.1924\pm0.0025$   	&	$4.96\pm0.24$   	&	 \\
7970740   	&	$10.67\pm0.36$   	&	$0.760\pm0.014$   	&	$0.7819\pm0.0055$   	&	$1.610\pm0.013$   	&	$0.444\pm0.029$   	&	 \\
8006161   	&	$4.75\pm0.25$   	&	$1.004\pm0.020$   	&	$0.9353\pm0.0070$   	&	$1.233\pm0.011$   	&	$0.717\pm0.040$   	&	 \\
8077137   	&	$4.48\pm0.62$   	&	$1.150\pm0.043$   	&	$1.638\pm0.027$   	&	$0.2626\pm0.0056$   	&	$3.37\pm0.25$   	&	2\footnotemark[6] \\
8150065   	&	$3.28\pm0.55$   	&	$1.196\pm0.066$   	&	$1.393\pm0.032$   	&	$0.445\pm0.011$   	&	$2.57\pm0.20$   	&	 \\
8179536   	&	$1.84\pm0.37$   	&	$1.248\pm0.059$   	&	$1.353\pm0.027$   	&	$0.5039\pm0.0081$   	&	$2.68\pm0.20$   	&	 \\
8228742   	&	$4.42\pm0.49$   	&	$1.221\pm0.069$   	&	$1.791\pm0.041$   	&	$0.2128\pm0.0030$   	&	$4.11\pm0.36$   	&	 \\
8292840   	&	$2.79\pm0.54$   	&	$1.099\pm0.036$   	&	$1.321\pm0.015$   	&	$0.4789\pm0.0060$   	&	$2.40\pm0.16$   	&	3\footnotemark[9]\footnotetext[9]{\citealt{2014ApJ...784...45R}} \\
\pagebreak 8349582   	&	$8.68\pm0.53$   	&	$1.144\pm0.015$   	&	$1.4520\pm0.0066$   	&	$0.3742\pm0.0072$   	&	$2.003\pm0.086$   	&	1\footnotemark[1] \Tstrut{}\\
8379927$^{\text{\hyperref[fig:scaling]{2}}}$   	&	$1.84\pm0.29$   	&	$1.105\pm0.049$   	&	$1.122\pm0.017$   	&	$0.790\pm0.012$   	&	$1.55\pm0.13$   	&	 \\
8394589   	&	$4.20\pm0.39$   	&	$1.024\pm0.037$   	&	$1.153\pm0.023$   	&	$0.673\pm0.019$   	&	$1.70\pm0.11$   	&	 \\
8424992   	&	$10.00\pm0.94$   	&	$0.893\pm0.038$   	&	$1.033\pm0.025$   	&	$0.819\pm0.027$   	&	$1.034\pm0.083$   	&	 \\
8478994$^{\text{\hyperref[fig:scaling]{9}}}$   	&	$5.68\pm0.96$   	&	$0.799\pm0.017$   	&	$0.776\pm0.014$   	&	$1.732\pm0.063$   	&	$0.482\pm0.032$   	&	4\footnotemark[10]$^,$\footnotetext[10]{\citealt{2014ApJ...787...80H}}\footnotemark[11]\footnotetext[11]{\citealt{2013Natur.494..452B}} \\
8494142   	&	$2.09\pm0.32$   	&	$1.458\pm0.056$   	&	$1.893\pm0.025$   	&	$0.2149\pm0.0060$   	&	$4.64\pm0.33$   	&	2\footnotemark[6] \\
8554498$^{\text{\hyperref[fig:scaling]{5}}}$   	&	$3.49\pm0.41$   	&	$1.320\pm0.026$   	&	$1.840\pm0.017$   	&	$0.2126\pm0.0030$   	&	$3.86\pm0.17$   	&	 \\
8684730   	&	$2.55\pm0.45$   	&	$1.475\pm0.039$   	&	$2.105\pm0.039$   	&	$0.163\pm0.011$   	&	$5.01\pm0.25$   	&	 \\
8694723$^{\text{\hyperref[fig:scaling]{11}}}$   	&	$5.26\pm0.23$   	&	$1.070\pm0.023$   	&	$1.515\pm0.013$   	&	$0.3084\pm0.0039$   	&	$3.16\pm0.19$   	&	 \\
8760414$^{\text{\hyperref[fig:scaling]{8}}}$   	&	$15.78\pm0.91$   	&	$0.761\pm0.011$   	&	$0.9755\pm0.0090$   	&	$0.828\pm0.015$   	&	$1.032\pm0.064$   	&	 \\
8866102   	&	$1.81\pm0.27$   	&	$1.223\pm0.054$   	&	$1.357\pm0.022$   	&	$0.4902\pm0.0048$   	&	$2.66\pm0.16$   	&	 \\
8938364   	&	$11.02\pm0.58$   	&	$0.995\pm0.023$   	&	$1.363\pm0.011$   	&	$0.3955\pm0.0046$   	&	$1.74\pm0.10$   	&	 \\
9025370$^{\text{\hyperref[fig:scaling]{1}}}$   	&	$5.77\pm0.60$   	&	$0.921\pm0.035$   	&	$0.987\pm0.019$   	&	$0.965\pm0.019$   	&	$0.709\pm0.084$   	&	 \\
9098294   	&	$9.42\pm0.85$   	&	$0.934\pm0.041$   	&	$1.112\pm0.027$   	&	$0.682\pm0.021$   	&	$1.301\pm0.081$   	&	 \\
9139151   	&	$1.82\pm0.19$   	&	$1.205\pm0.044$   	&	$1.168\pm0.016$   	&	$0.7566\pm0.0097$   	&	$1.96\pm0.12$   	&	 \\
9139163   	&	$2.09\pm0.28$   	&	$1.378\pm0.039$   	&	$1.557\pm0.017$   	&	$0.3653\pm0.0038$   	&	$3.70\pm0.22$   	&	 \\
9206432$^{\text{\hyperref[fig:scaling]{12}}}$   	&	$1.67\pm0.42$   	&	$1.355\pm0.047$   	&	$1.505\pm0.021$   	&	$0.3982\pm0.0065$   	&	$3.76\pm0.21$   	&	 \\
9353712   	&	$2.44\pm0.21$   	&	$1.465\pm0.054$   	&	$2.164\pm0.032$   	&	$0.1446\pm0.0019$   	&	$6.60\pm0.50$   	&	 \\
9410862   	&	$7.34\pm0.75$   	&	$0.954\pm0.028$   	&	$1.139\pm0.017$   	&	$0.650\pm0.019$   	&	$1.580\pm0.099$   	&	 \\
9414417   	&	$2.80\pm0.22$   	&	$1.319\pm0.041$   	&	$1.882\pm0.023$   	&	$0.1978\pm0.0026$   	&	$4.89\pm0.29$   	&	 \\
9592705   	&	$2.21\pm0.19$   	&	$1.521\pm0.044$   	&	$2.118\pm0.026$   	&	$0.1610\pm0.0036$   	&	$5.94\pm0.37$   	&	 \\
9812850   	&	$1.99\pm0.39$   	&	$1.417\pm0.057$   	&	$1.805\pm0.025$   	&	$0.2399\pm0.0067$   	&	$4.71\pm0.24$   	&	 \\
9955598   	&	$6.69\pm0.48$   	&	$0.921\pm0.032$   	&	$0.893\pm0.012$   	&	$1.306\pm0.015$   	&	$0.634\pm0.042$   	&	1\footnotemark[1] \\
9965715$^{\text{\hyperref[fig:scaling]{6}}}$   	&	$3.88\pm0.49$   	&	$1.030\pm0.023$   	&	$1.244\pm0.011$   	&	$0.538\pm0.013$   	&	$1.78\pm0.17$   	&	 \\
10068307$^{\text{\hyperref[fig:scaling]{7}}}$   	&	$3.98\pm0.34$   	&	$1.271\pm0.045$   	&	$2.024\pm0.023$   	&	$0.1546\pm0.0015$   	&	$5.24\pm0.37$   	&	 \\
10079226   	&	$2.56\pm0.40$   	&	$1.153\pm0.050$   	&	$1.160\pm0.016$   	&	$0.744\pm0.013$   	&	$1.521\pm0.088$   	&	 \\
10162436   	&	$3.29\pm0.42$   	&	$1.379\pm0.052$   	&	$2.019\pm0.026$   	&	$0.1678\pm0.0017$   	&	$5.25\pm0.31$   	&	 \\
10454113$^{\text{\hyperref[fig:scaling]{3}}}$   	&	$5.07\pm0.91$   	&	$1.078\pm0.048$   	&	$1.202\pm0.022$   	&	$0.659\pm0.018$   	&	$1.94\pm0.13$   	&	 \\
10514430   	&	$7.01\pm0.61$   	&	$1.042\pm0.029$   	&	$1.578\pm0.016$   	&	$0.2664\pm0.0034$   	&	$2.55\pm0.16$   	&	 \\
10516096   	&	$6.66\pm0.47$   	&	$1.108\pm0.035$   	&	$1.422\pm0.017$   	&	$0.3861\pm0.0029$   	&	$2.30\pm0.17$   	&	 \\
10586004   	&	$6.13\pm0.89$   	&	$1.154\pm0.041$   	&	$1.643\pm0.029$   	&	$0.2605\pm0.0056$   	&	$2.73\pm0.20$   	&	2\footnotemark[9] \\
10644253   	&	$1.28\pm0.25$   	&	$1.128\pm0.027$   	&	$1.1204\pm0.0085$   	&	$0.804\pm0.012$   	&	$1.525\pm0.060$   	&	 \\
10666592   	&	$2.02\pm0.25$   	&	$1.556\pm0.046$   	&	$2.009\pm0.023$   	&	$0.1918\pm0.0027$   	&	$5.92\pm0.42$   	&	1\footnotemark[12]\footnotetext[12]{\citealt{2008ApJ...680.1450P}} \\
10730618   	&	$4.4\pm1.1$   	&	$1.249\pm0.085$   	&	$1.716\pm0.053$   	&	$0.251\pm0.012$   	&	$3.81\pm0.52$   	&	 \\
10963065   	&	$4.57\pm0.48$   	&	$1.065\pm0.043$   	&	$1.225\pm0.026$   	&	$0.584\pm0.017$   	&	$1.93\pm0.13$   	&	1\footnotemark[1] \\
11081729   	&	$1.66\pm0.84$   	&	$1.363\pm0.039$   	&	$1.444\pm0.018$   	&	$0.4526\pm0.0082$   	&	$3.46\pm0.20$   	&	 \\
11133306   	&	$4.58\pm0.74$   	&	$1.081\pm0.040$   	&	$1.198\pm0.016$   	&	$0.634\pm0.015$   	&	$1.66\pm0.11$   	&	 \\
11253226   	&	$1.34\pm0.50$   	&	$1.420\pm0.071$   	&	$1.599\pm0.026$   	&	$0.3468\pm0.0094$   	&	$4.53\pm0.29$   	&	 \\
11295426   	&	$6.31\pm0.39$   	&	$1.099\pm0.031$   	&	$1.251\pm0.013$   	&	$0.5673\pm0.0080$   	&	$1.587\pm0.096$   	&	3\footnotemark[13]\footnotetext[13]{\citealt{2013ApJ...766...40G}} \\
11401755   	&	$6.94\pm0.78$   	&	$1.043\pm0.031$   	&	$1.601\pm0.026$   	&	$0.2552\pm0.0074$   	&	$2.91\pm0.21$   	&	2\footnotemark[14]\footnotetext[14]{\citealt{2012Sci...337..556C}} \\
11772920   	&	$9.9\pm1.1$   	&	$0.823\pm0.040$   	&	$0.845\pm0.018$   	&	$1.373\pm0.024$   	&	$0.468\pm0.068$   	&	 \\
11807274   	&	$2.01\pm0.28$   	&	$1.306\pm0.032$   	&	$1.597\pm0.013$   	&	$0.3223\pm0.0052$   	&	$3.51\pm0.17$   	&	2\footnotemark[15]\footnotetext[15]{\citealt{2013MNRAS.428.1077S}} \\
11853905   	&	$6.60\pm0.57$   	&	$1.163\pm0.046$   	&	$1.588\pm0.019$   	&	$0.2911\pm0.0042$   	&	$2.53\pm0.17$   	&	1\footnotemark[16]\footnotetext[16]{\citealt{2010ApJ...713L.126B}} \\
11904151   	&	$11.65\pm0.93$   	&	$0.880\pm0.031$   	&	$1.041\pm0.016$   	&	$0.784\pm0.022$   	&	$0.995\pm0.054$   	&	2\footnotemark[17]$^,$\footnotetext[17]{\citealt{2011ApJS..197....5F}}\footnotemark[18]\footnotetext[18]{\citealt{2011ApJ...729...27B}} \\
12009504   	&	$3.80\pm0.29$   	&	$1.197\pm0.044$   	&	$1.411\pm0.018$   	&	$0.4266\pm0.0054$   	&	$2.63\pm0.16$   	&	 \\
12069127   	&	$2.31\pm0.18$   	&	$1.537\pm0.051$   	&	$2.297\pm0.030$   	&	$0.1267\pm0.0015$   	&	$7.48\pm0.44$   	&	 \\
12069424   	&	$6.92\pm0.25$   	&	$1.056\pm0.019$   	&	$1.213\pm0.011$   	&	$0.5967\pm0.0085$   	&	$1.531\pm0.076$   	&	 \\
12069449   	&	$7.08\pm0.22$   	&	$1.0000\pm0.0086$   	&	$1.101\pm0.012$   	&	$0.758\pm0.018$   	&	$1.201\pm0.024$   	&	1\footnotemark[19]\footnotetext[19]{\citealt{1997ApJ...483..457C}} \\
12258514   	&	$4.40\pm0.29$   	&	$1.252\pm0.054$   	&	$1.614\pm0.023$   	&	$0.2999\pm0.0024$   	&	$2.95\pm0.18$   	&	 \\
12317678   	&	$2.63\pm0.28$   	&	$1.288\pm0.036$   	&	$1.775\pm0.022$   	&	$0.2301\pm0.0042$   	&	$5.33\pm0.25$   	&	\\\hline
    %\end{tabular}
\end{longtable}
%\end{center}
%-------------------------------------------------------------------
%\clearpage
%\twocolumn
\begin{multicols}{2}
\bibliographystyle{aa.bst}
\bibliography{aa}

\begin{thebibliography}{111}
\expandafter\ifx\csname natexlab\endcsname\relax\def\natexlab#1{#1}\fi

\bibitem[{{Aarslev} {et~al.}(2017){Aarslev}, {Christensen-Dalsgaard}, {Lund},
  {Silva Aguirre}, \& {Gough}}]{2017EPJWC.16003010A}
{Aarslev}, M.~J., {Christensen-Dalsgaard}, J., {Lund}, M.~N., {Silva Aguirre},
  V., \& {Gough}, D. 2017, in European Physical Journal Web of Conferences,
  Vol. 160, European Physical Journal Web of Conferences, 03010

\bibitem[{{Adibekyan} {et~al.}(2018){Adibekyan}, {Sousa}, \&
  {Santos}}]{2018ASSP...49..225A}
{Adibekyan}, V., {Sousa}, S.~G., \& {Santos}, N.~C. 2018, Asteroseismology and
  Exoplanets: Listening to the Stars and Searching for New Worlds, 49, 225

\bibitem[{{Akeson} {et~al.}(2013){Akeson}, {Chen}, {Ciardi}, {Crane}, {Good},
  {Harbut}, {Jackson}, {Kane}, {Laity}, {Leifer}, {Lynn}, {McElroy}, {Papin},
  {Plavchan}, {Ram{\'{\i}}rez}, {Rey}, {von Braun}, {Wittman}, {Abajian},
  {Ali}, {Beichman}, {Beekley}, {Berriman}, {Berukoff}, {Bryden}, {Chan},
  {Groom}, {Lau}, {Payne}, {Regelson}, {Saucedo}, {Schmitz}, {Stauffer},
  {Wyatt}, \& {Zhang}}]{2013PASP..125..989A}
{Akeson}, R.~L., {Chen}, X., {Ciardi}, D., {et~al.} 2013, \pasp, 125, 989

\bibitem[{{Anders} {et~al.}(2017){Anders}, {Chiappini}, {Rodrigues}, {Miglio},
  {Montalb{\'a}n}, {Mosser}, {Girardi}, {Valentini}, {Noels}, {Morel},
  {Johnson}, {Schultheis}, {Baudin}, {de Assis Peralta}, {Hekker},
  {Theme{\ss}l}, {Kallinger}, {Garc{\'{\i}}a}, {Mathur}, {Baglin}, {Santiago},
  {Martig}, {Minchev}, {Steinmetz}, {da Costa}, {Maia}, {Allende Prieto},
  {Cunha}, {Beers}, {Epstein}, {Garc{\'{\i}}a P{\'e}rez},
  {Garc{\'{\i}}a-Hern{\'a}ndez}, {Harding}, {Holtzman}, {Majewski},
  {M{\'e}sz{\'a}ros}, {Nidever}, {Pan}, {Pinsonneault}, {Schiavon},
  {Schneider}, {Shetrone}, {Stassun}, {Zamora}, \&
  {Zasowski}}]{2017A&A...597A..30A}
{Anders}, F., {Chiappini}, C., {Rodrigues}, T.~S., {et~al.} 2017, \aap, 597,
  A30

\bibitem[{{Andrae} {et~al.}(2018){Andrae}, {Fouesneau}, {Creevey}, {Ordenovic},
  {Mary}, {Burlacu}, {Chaoul}, {Jean-Antoine-Piccolo}, {Kordopatis}, {Korn},
  {Lebreton}, {Panem}, {Pichon}, {Th{\'e}venin}, {Walmsley}, \&
  {Bailer-Jones}}]{2018A&A...616A...8A}
{Andrae}, R., {Fouesneau}, M., {Creevey}, O., {et~al.} 2018, \aap, 616, A8

\bibitem[{{Angelou} {et~al.}(2017){Angelou}, {Bellinger}, {Hekker}, \&
  {Basu}}]{2017ApJ...839..116A}
{Angelou}, G.~C., {Bellinger}, E.~P., {Hekker}, S., \& {Basu}, S. 2017, \apj,
  839, 116

\bibitem[{{Barclay} {et~al.}(2013){Barclay}, {Rowe}, {Lissauer}, {Huber},
  {Fressin}, {Howell}, {Bryson}, {Chaplin}, {D{\'e}sert}, {Lopez}, {Marcy},
  {Mullally}, {Ragozzine}, {Torres}, {Adams}, {Agol}, {Barrado}, {Basu},
  {Bedding}, {Buchhave}, {Charbonneau}, {Christiansen},
  {Christensen-Dalsgaard}, {Ciardi}, {Cochran}, {Dupree}, {Elsworth},
  {Everett}, {Fischer}, {Ford}, {Fortney}, {Geary}, {Haas}, {Handberg},
  {Hekker}, {Henze}, {Horch}, {Howard}, {Hunter}, {Isaacson}, {Jenkins},
  {Karoff}, {Kawaler}, {Kjeldsen}, {Klaus}, {Latham}, {Li}, {Lillo-Box},
  {Lund}, {Lundkvist}, {Metcalfe}, {Miglio}, {Morris}, {Quintana}, {Stello},
  {Smith}, {Still}, \& {Thompson}}]{2013Natur.494..452B}
{Barclay}, T., {Rowe}, J.~F., {Lissauer}, J.~J., {et~al.} 2013, \nat, 494, 452

\bibitem[{{Basu}(1997)}]{1997MNRAS.288..572B}
{Basu}, S. 1997, \mnras, 288, 572

\bibitem[{Basu \& Chaplin(2017)}]{basuchaplin2017}
Basu, S. \& Chaplin, W. 2017, Asteroseismic Data Analysis: Foundations and
  Techniques, Princeton Series in Modern Obs (Princeton University Press)

\bibitem[{{Basu} \& {Kinnane}(2018)}]{BasuKinnane2018}
{Basu}, S. \& {Kinnane}, A. 2018, ArXiv e-prints [\eprint[arXiv]{1810.07205}]

\bibitem[{{Batalha} {et~al.}(2011){Batalha}, {Borucki}, {Bryson}, {Buchhave},
  {Caldwell}, {Christensen-Dalsgaard}, {Ciardi}, {Dunham}, {Fressin},
  {Gautier}, {Gilliland}, {Haas}, {Howell}, {Jenkins}, {Kjeldsen}, {Koch},
  {Latham}, {Lissauer}, {Marcy}, {Rowe}, {Sasselov}, {Seager}, {Steffen},
  {Torres}, {Basri}, {Brown}, {Charbonneau}, {Christiansen}, {Clarke},
  {Cochran}, {Dupree}, {Fabrycky}, {Fischer}, {Ford}, {Fortney}, {Girouard},
  {Holman}, {Johnson}, {Isaacson}, {Klaus}, {Machalek}, {Moorehead},
  {Morehead}, {Ragozzine}, {Tenenbaum}, {Twicken}, {Quinn}, {VanCleve},
  {Walkowicz}, {Welsh}, {Devore}, \& {Gould}}]{2011ApJ...729...27B}
{Batalha}, N.~M., {Borucki}, W.~J., {Bryson}, S.~T., {et~al.} 2011, \apj, 729,
  27

\bibitem[{{Belkacem} {et~al.}(2011){Belkacem}, {Goupil}, {Dupret}, {Samadi},
  {Baudin}, {Noels}, \& {Mosser}}]{2011A&A...530A.142B}
{Belkacem}, K., {Goupil}, M.~J., {Dupret}, M.~A., {et~al.} 2011, \aap, 530,
  A142

\bibitem[{{Bellinger} {et~al.}(2016){Bellinger}, {Angelou}, {Hekker}, {Basu},
  {Ball}, \& {Guggenberger}}]{2016ApJ...830...31B}
{Bellinger}, E.~P., {Angelou}, G.~C., {Hekker}, S., {et~al.} 2016, \apj, 830,
  31

\bibitem[{{Bellinger} {et~al.}(2017){Bellinger}, {Basu}, {Hekker}, \&
  {Ball}}]{2017ApJ...851...80B}
{Bellinger}, E.~P., {Basu}, S., {Hekker}, S., \& {Ball}, W.~H. 2017, \apj, 851,
  80

\bibitem[{{Borucki} {et~al.}(2010{\natexlab{a}}){Borucki}, {Koch}, {Basri},
  {Batalha}, {Brown}, {Caldwell}, {Caldwell}, {Christensen-Dalsgaard},
  {Cochran}, {DeVore}, {Dunham}, {Dupree}, {Gautier}, {Geary}, {Gilliland},
  {Gould}, {Howell}, {Jenkins}, {Kondo}, {Latham}, {Marcy}, {Meibom},
  {Kjeldsen}, {Lissauer}, {Monet}, {Morrison}, {Sasselov}, {Tarter}, {Boss},
  {Brownlee}, {Owen}, {Buzasi}, {Charbonneau}, {Doyle}, {Fortney}, {Ford},
  {Holman}, {Seager}, {Steffen}, {Welsh}, {Rowe}, {Anderson}, {Buchhave},
  {Ciardi}, {Walkowicz}, {Sherry}, {Horch}, {Isaacson}, {Everett}, {Fischer},
  {Torres}, {Johnson}, {Endl}, {MacQueen}, {Bryson}, {Dotson}, {Haas},
  {Kolodziejczak}, {Van Cleve}, {Chandrasekaran}, {Twicken}, {Quintana},
  {Clarke}, {Allen}, {Li}, {Wu}, {Tenenbaum}, {Verner}, {Bruhweiler}, {Barnes},
  \& {Prsa}}]{2010Sci...327..977B}
{Borucki}, W.~J., {Koch}, D., {Basri}, G., {et~al.} 2010{\natexlab{a}},
  Science, 327, 977

\bibitem[{{Borucki} {et~al.}(2010{\natexlab{b}}){Borucki}, {Koch}, {Brown},
  {Basri}, {Batalha}, {Caldwell}, {Cochran}, {Dunham}, {Gautier}, {Geary},
  {Gilliland}, {Howell}, {Jenkins}, {Latham}, {Lissauer}, {Marcy}, {Monet},
  {Rowe}, \& {Sasselov}}]{2010ApJ...713L.126B}
{Borucki}, W.~J., {Koch}, D.~G., {Brown}, T.~M., {et~al.} 2010{\natexlab{b}},
  \apjl, 713, L126

\bibitem[{Breiman(2001)}]{breiman2001random}
Breiman, L. 2001, Machine Learning, 45, 5

\bibitem[{{Brown} {et~al.}(1994){Brown}, {Christensen-Dalsgaard},
  {Weibel-Mihalas}, \& {Gilliland}}]{1994ApJ...427.1013B}
{Brown}, T.~M., {Christensen-Dalsgaard}, J., {Weibel-Mihalas}, B., \&
  {Gilliland}, R.~L. 1994, \apj, 427, 1013

\bibitem[{{Brown} {et~al.}(1991){Brown}, {Gilliland}, {Noyes}, \&
  {Ramsey}}]{1991ApJ...368..599B}
{Brown}, T.~M., {Gilliland}, R.~L., {Noyes}, R.~W., \& {Ramsey}, L.~W. 1991,
  \apj, 368, 599

\bibitem[{{Campante} {et~al.}(2015){Campante}, {Barclay}, {Swift}, {Huber},
  {Adibekyan}, {Cochran}, {Burke}, {Isaacson}, {Quintana}, {Davies}, {Silva
  Aguirre}, {Ragozzine}, {Riddle}, {Baranec}, {Basu}, {Chaplin},
  {Christensen-Dalsgaard}, {Metcalfe}, {Bedding}, {Handberg}, {Stello},
  {Brewer}, {Hekker}, {Karoff}, {Kolbl}, {Law}, {Lundkvist}, {Miglio}, {Rowe},
  {Santos}, {Van Laerhoven}, {Arentoft}, {Elsworth}, {Fischer}, {Kawaler},
  {Kjeldsen}, {Lund}, {Marcy}, {Sousa}, {Sozzetti}, \&
  {White}}]{2015ApJ...799..170C}
{Campante}, T.~L., {Barclay}, T., {Swift}, J.~J., {et~al.} 2015, \apj, 799, 170

\bibitem[{{Campante} {et~al.}(2016){Campante}, {Lund}, {Kuszlewicz}, {Davies},
  {Chaplin}, {Albrecht}, {Winn}, {Bedding}, {Benomar}, {Bossini}, {Handberg},
  {Santos}, {Van Eylen}, {Basu}, {Christensen-Dalsgaard}, {Elsworth}, {Hekker},
  {Hirano}, {Huber}, {Karoff}, {Kjeldsen}, {Lundkvist}, {North}, {Silva
  Aguirre}, {Stello}, \& {White}}]{2016ApJ...819...85C}
{Campante}, T.~L., {Lund}, M.~N., {Kuszlewicz}, J.~S., {et~al.} 2016, \apj,
  819, 85

\bibitem[{{Carter} {et~al.}(2012){Carter}, {Agol}, {Chaplin}, {Basu},
  {Bedding}, {Buchhave}, {Christensen-Dalsgaard}, {Deck}, {Elsworth},
  {Fabrycky}, {Ford}, {Fortney}, {Hale}, {Handberg}, {Hekker}, {Holman},
  {Huber}, {Karoff}, {Kawaler}, {Kjeldsen}, {Lissauer}, {Lopez}, {Lund},
  {Lundkvist}, {Metcalfe}, {Miglio}, {Rogers}, {Stello}, {Borucki}, {Bryson},
  {Christiansen}, {Cochran}, {Geary}, {Gilliland}, {Haas}, {Hall}, {Howard},
  {Jenkins}, {Klaus}, {Koch}, {Latham}, {MacQueen}, {Sasselov}, {Steffen},
  {Twicken}, \& {Winn}}]{2012Sci...337..556C}
{Carter}, J.~A., {Agol}, E., {Chaplin}, W.~J., {et~al.} 2012, Science, 337, 556

\bibitem[{{Chaplin} \& {Miglio}(2013)}]{2013ARA&A..51..353C}
{Chaplin}, W.~J. \& {Miglio}, A. 2013, \araa, 51, 353

\bibitem[{{Chaplin} {et~al.}(2013){Chaplin}, {Sanchis-Ojeda}, {Campante},
  {Handberg}, {Stello}, {Winn}, {Basu}, {Christensen-Dalsgaard}, {Davies},
  {Metcalfe}, {Buchhave}, {Fischer}, {Bedding}, {Cochran}, {Elsworth},
  {Gilliland}, {Hekker}, {Huber}, {Isaacson}, {Karoff}, {Kawaler}, {Kjeldsen},
  {Latham}, {Lund}, {Lundkvist}, {Marcy}, {Miglio}, {Barclay}, \&
  {Lissauer}}]{2013ApJ...766..101C}
{Chaplin}, W.~J., {Sanchis-Ojeda}, R., {Campante}, T.~L., {et~al.} 2013, \apj,
  766, 101

\bibitem[{{Chiappini} {et~al.}(2015){Chiappini}, {Anders}, {Rodrigues},
  {Miglio}, {Montalb{\'a}n}, {Mosser}, {Girardi}, {Valentini}, {Noels},
  {Morel}, {Minchev}, {Steinmetz}, {Santiago}, {Schultheis}, {Martig}, {da
  Costa}, {Maia}, {Allende Prieto}, {de Assis Peralta}, {Hekker},
  {Theme{\ss}l}, {Kallinger}, {Garc{\'{\i}}a}, {Mathur}, {Baudin}, {Beers},
  {Cunha}, {Harding}, {Holtzman}, {Majewski}, {M{\'e}sz{\'a}ros}, {Nidever},
  {Pan}, {Schiavon}, {Shetrone}, {Schneider}, \&
  {Stassun}}]{2015A&A...576L..12C}
{Chiappini}, C., {Anders}, F., {Rodrigues}, T.~S., {et~al.} 2015, \aap, 576,
  L12

\bibitem[{{Christensen-Dalsgaard}(1984)}]{1984srps.conf...11C}
{Christensen-Dalsgaard}, J. 1984, in Space Research in Stellar Activity and
  Variability, ed. A.~{Mangeney} \& F.~{Praderie}, 11

\bibitem[{{Christensen-Dalsgaard} \& {Silva
  Aguirre}(2018)}]{2018arXiv180303125C}
{Christensen-Dalsgaard}, J. \& {Silva Aguirre}, V. 2018, in {Handbook of
  Exoplanets}, ed. {{Deeg}, Hans J. and {Belmonte}, Juan Antonio} ({Springer}),
  18

\bibitem[{{Cochran} {et~al.}(1997){Cochran}, {Hatzes}, {Butler}, \&
  {Marcy}}]{1997ApJ...483..457C}
{Cochran}, W.~D., {Hatzes}, A.~P., {Butler}, R.~P., \& {Marcy}, G.~W. 1997,
  \apj, 483, 457

\bibitem[{{Creevey} {et~al.}(2013){Creevey}, {Th{\'e}venin}, {Basu}, {Chaplin},
  {Bigot}, {Elsworth}, {Huber}, {Monteiro}, \&
  {Serenelli}}]{2013MNRAS.431.2419C}
{Creevey}, O.~L., {Th{\'e}venin}, F., {Basu}, S., {et~al.} 2013, \mnras, 431,
  2419

\bibitem[{{Davies} {et~al.}(2016){Davies}, {Silva Aguirre}, {Bedding},
  {Handberg}, {Lund}, {Chaplin}, {Huber}, {White}, {Benomar}, {Hekker}, {Basu},
  {Campante}, {Christensen-Dalsgaard}, {Elsworth}, {Karoff}, {Kjeldsen},
  {Lundkvist}, {Metcalfe}, \& {Stello}}]{2016MNRAS.456.2183D}
{Davies}, G.~R., {Silva Aguirre}, V., {Bedding}, T.~R., {et~al.} 2016, \mnras,
  456, 2183

\bibitem[{{Deheuvels} {et~al.}(2016){Deheuvels}, {Brand{\~a}o}, {Silva
  Aguirre}, {Ballot}, {Michel}, {Cunha}, {Lebreton}, \&
  {Appourchaux}}]{2016A&A...589A..93D}
{Deheuvels}, S., {Brand{\~a}o}, I., {Silva Aguirre}, V., {et~al.} 2016, \aap,
  589, A93

\bibitem[{{Feuillet} {et~al.}(2018){Feuillet}, {Bovy}, {Holtzman}, {Weinberg},
  {Garc{\'{\i}}a-Hern{\'a}ndez}, {Hearty}, {Majewski}, {Roman-Lopes},
  {Rybizki}, \& {Zamora}}]{2018MNRAS.477.2326F}
{Feuillet}, D.~K., {Bovy}, J., {Holtzman}, J., {et~al.} 2018, \mnras, 477, 2326

\bibitem[{{Fressin} {et~al.}(2011){Fressin}, {Torres}, {D{\'e}sert},
  {Charbonneau}, {Batalha}, {Fortney}, {Rowe}, {Allen}, {Borucki}, {Brown},
  {Bryson}, {Ciardi}, {Cochran}, {Deming}, {Dunham}, {Fabrycky}, {Gautier},
  {Gilliland}, {Henze}, {Holman}, {Howell}, {Jenkins}, {Kinemuchi}, {Knutson},
  {Koch}, {Latham}, {Lissauer}, {Marcy}, {Ragozzine}, {Sasselov}, {Still},
  {Tenenbaum}, \& {Uddin}}]{2011ApJS..197....5F}
{Fressin}, F., {Torres}, G., {D{\'e}sert}, J.-M., {et~al.} 2011, \apjs, 197, 5

\bibitem[{Friedman {et~al.}(2001)Friedman, Hastie, \&
  Tibshirani}]{friedman2001elements}
Friedman, J., Hastie, T., \& Tibshirani, R. 2001, The Elements of Statistical
  Learning, Vol.~1 (Springer New York)

\bibitem[{{Gai} {et~al.}(2011){Gai}, {Basu}, {Chaplin}, \&
  {Elsworth}}]{2011ApJ...730...63G}
{Gai}, N., {Basu}, S., {Chaplin}, W.~J., \& {Elsworth}, Y. 2011, \apj, 730, 63

\bibitem[{{Gaia Collaboration} {et~al.}(2018){Gaia Collaboration}, {Brown},
  {Vallenari}, {Prusti}, {de Bruijne}, {Babusiaux}, {Bailer-Jones}, {Biermann},
  {Evans}, {Eyer}, \& et~al.}]{2018A&A...616A...1G}
{Gaia Collaboration}, {Brown}, A.~G.~A., {Vallenari}, A., {et~al.} 2018, \aap,
  616, A1

\bibitem[{{Gaulme} {et~al.}(2016){Gaulme}, {McKeever}, {Jackiewicz}, {Rawls},
  {Corsaro}, {Mosser}, {Southworth}, {Mahadevan}, {Bender}, \&
  {Deshpande}}]{2016ApJ...832..121G}
{Gaulme}, P., {McKeever}, J., {Jackiewicz}, J., {et~al.} 2016, \apj, 832, 121

\bibitem[{Geurts {et~al.}(2006)Geurts, Ernst, \&
  Wehenkel}]{geurts2006extremely}
Geurts, P., Ernst, D., \& Wehenkel, L. 2006, Machine Learning, 63, 3

\bibitem[{{Gilliland} {et~al.}(2013){Gilliland}, {Marcy}, {Rowe}, {Rogers},
  {Torres}, {Fressin}, {Lopez}, {Buchhave}, {Christensen-Dalsgaard},
  {D{\'e}sert}, {Henze}, {Isaacson}, {Jenkins}, {Lissauer}, {Chaplin}, {Basu},
  {Metcalfe}, {Elsworth}, {Handberg}, {Hekker}, {Huber}, {Karoff}, {Kjeldsen},
  {Lund}, {Lundkvist}, {Miglio}, {Charbonneau}, {Ford}, {Fortney}, {Haas},
  {Howard}, {Howell}, {Ragozzine}, \& {Thompson}}]{2013ApJ...766...40G}
{Gilliland}, R.~L., {Marcy}, G.~W., {Rowe}, J.~F., {et~al.} 2013, \apj, 766, 40

\bibitem[{Ginsburg {et~al.}(2018)Ginsburg, Sipocz, Parikh, Woillez, Groener,
  Liedtke, Robitaille, Deil, jcsegovia, Norman, Svoboda, Brasseur, Tollerud,
  Persson, adamginsburg, Séguin-Charbonneau, Armstrong, de~Val-Borro, Morris,
  Mirocha, Yadav, Seifert, Droettboom, Moolekamp, james allen, Bostroem,
  Egeland, Singer, Rol, \& Grollier}]{adam_ginsburg_2018_1160627}
Ginsburg, A., Sipocz, B., Parikh, M., {et~al.} 2018, astropy/astroquery: v0.3.7
  release

\bibitem[{{Gratia} \& {Fabrycky}(2017)}]{2017MNRAS.464.1709G}
{Gratia}, P. \& {Fabrycky}, D. 2017, \mnras, 464, 1709

\bibitem[{{Guggenberger} {et~al.}(2017){Guggenberger}, {Hekker}, {Angelou},
  {Basu}, \& {Bellinger}}]{2017MNRAS.470.2069G}
{Guggenberger}, E., {Hekker}, S., {Angelou}, G.~C., {Basu}, S., \& {Bellinger},
  E.~P. 2017, \mnras, 470, 2069

\bibitem[{{Guggenberger} {et~al.}(2016){Guggenberger}, {Hekker}, {Basu}, \&
  {Bellinger}}]{2016MNRAS.460.4277G}
{Guggenberger}, E., {Hekker}, S., {Basu}, S., \& {Bellinger}, E. 2016, \mnras,
  460, 4277

\bibitem[{{Guillochon} {et~al.}(2011){Guillochon}, {Ramirez-Ruiz}, \&
  {Lin}}]{2011ApJ...732...74G}
{Guillochon}, J., {Ramirez-Ruiz}, E., \& {Lin}, D. 2011, \apj, 732, 74

\bibitem[{{Hadden} \& {Lithwick}(2014)}]{2014ApJ...787...80H}
{Hadden}, S. \& {Lithwick}, Y. 2014, \apj, 787, 80

\bibitem[{{Han} {et~al.}(2014){Han}, {Wang}, {Wright}, {Feng}, {Zhao},
  {Fakhouri}, {Brown}, \& {Hancock}}]{2014PASP..126..827H}
{Han}, E., {Wang}, S.~X., {Wright}, J.~T., {et~al.} 2014, \pasp, 126, 827

\bibitem[{{Haywood} {et~al.}(2018){Haywood}, {Vanderburg}, {Mortier}, {Giles},
  {L{\'o}pez-Morales}, {Lopez}, {Malavolta}, {Charbonneau}, {Collier Cameron},
  {Coughlin}, {Dressing}, {Nava}, {Latham}, {Dumusque}, {Lovis}, {Molinari},
  {Pepe}, {Sozzetti}, {Udry}, {Bouchy}, {Johnson}, {Mayor}, {Micela},
  {Phillips}, {Piotto}, {Rice}, {Sasselov}, {S{\'e}gransan}, {Watson}, {Affer},
  {Bonomo}, {Buchhave}, {Ciardi}, {Fiorenzano}, \&
  {Harutyunyan}}]{2018AJ....155..203H}
{Haywood}, R.~D., {Vanderburg}, A., {Mortier}, A., {et~al.} 2018, \aj, 155, 203

\bibitem[{{Hekker} {et~al.}(2013){Hekker}, {Elsworth}, {Mosser}, {Kallinger},
  {Basu}, {Chaplin}, \& {Stello}}]{2013A&A...556A..59H}
{Hekker}, S., {Elsworth}, Y., {Mosser}, B., {et~al.} 2013, \aap, 556, A59

\bibitem[{{Hj{\o}rringgaard} {et~al.}(2017){Hj{\o}rringgaard}, {Silva Aguirre},
  {White}, {Huber}, {Pope}, {Casagrande}, {Justesen}, \&
  {Christensen-Dalsgaard}}]{2017MNRAS.464.3713H}
{Hj{\o}rringgaard}, J.~G., {Silva Aguirre}, V., {White}, T.~R., {et~al.} 2017,
  \mnras, 464, 3713

\bibitem[{{Howell} {et~al.}(2012){Howell}, {Rowe}, {Bryson}, {Quinn}, {Marcy},
  {Isaacson}, {Ciardi}, {Chaplin}, {Metcalfe}, {Monteiro}, {Appourchaux},
  {Basu}, {Creevey}, {Gilliland}, {Quirion}, {Stello}, {Kjeldsen},
  {Christensen-Dalsgaard}, {Elsworth}, {Garc{\'{\i}}a}, {Houdek}, {Karoff},
  {Molenda-{\.Z}akowicz}, {Thompson}, {Verner}, {Torres}, {Fressin}, {Crepp},
  {Adams}, {Dupree}, {Sasselov}, {Dressing}, {Borucki}, {Koch}, {Lissauer},
  {Latham}, {Buchhave}, {Gautier}, {Everett}, {Horch}, {Batalha}, {Dunham},
  {Szkody}, {Silva}, {Mighell}, {Holberg}, {Ballot}, {Bedding}, {Bruntt},
  {Campante}, {Handberg}, {Hekker}, {Huber}, {Mathur}, {Mosser}, {R{\'e}gulo},
  {White}, {Christiansen}, {Middour}, {Haas}, {Hall}, {Jenkins}, {McCaulif},
  {Fanelli}, {Kulesa}, {McCarthy}, \& {Henze}}]{2012ApJ...746..123H}
{Howell}, S.~B., {Rowe}, J.~F., {Bryson}, S.~T., {et~al.} 2012, \apj, 746, 123

\bibitem[{{Huber}(2016)}]{2016IAUFM..29B.620H}
{Huber}, D. 2016, IAU Focus Meeting, 29, 620

\bibitem[{{Huber}(2018)}]{2018ASSP...49..119H}
{Huber}, D. 2018, Asteroseismology and Exoplanets: Listening to the Stars and
  Searching for New Worlds, 49, 119

\bibitem[{{Huber} {et~al.}(2011){Huber}, {Bedding}, {Stello}, {Hekker},
  {Mathur}, {Mosser}, {Verner}, {Bonanno}, {Buzasi}, {Campante}, {Elsworth},
  {Hale}, {Kallinger}, {Silva Aguirre}, {Chaplin}, {De Ridder},
  {Garc{\'{\i}}a}, {Appourchaux}, {Frandsen}, {Houdek}, {Molenda-{\.Z}akowicz},
  {Monteiro}, {Christensen-Dalsgaard}, {Gilliland}, {Kawaler}, {Kjeldsen},
  {Broomhall}, {Corsaro}, {Salabert}, {Sanderfer}, {Seader}, \&
  {Smith}}]{2011ApJ...743..143H}
{Huber}, D., {Bedding}, T.~R., {Stello}, D., {et~al.} 2011, \apj, 743, 143

\bibitem[{{Huber} {et~al.}(2013){Huber}, {Carter}, {Barbieri}, {Miglio},
  {Deck}, {Fabrycky}, {Montet}, {Buchhave}, {Chaplin}, {Hekker},
  {Montalb{\'a}n}, {Sanchis-Ojeda}, {Basu}, {Bedding}, {Campante},
  {Christensen-Dalsgaard}, {Elsworth}, {Stello}, {Arentoft}, {Ford},
  {Gilliland}, {Handberg}, {Howard}, {Isaacson}, {Johnson}, {Karoff},
  {Kawaler}, {Kjeldsen}, {Latham}, {Lund}, {Lundkvist}, {Marcy}, {Metcalfe},
  {Silva Aguirre}, \& {Winn}}]{2013Sci...342..331H}
{Huber}, D., {Carter}, J.~A., {Barbieri}, M., {et~al.} 2013, Science, 342, 331

\bibitem[{{Huber} {et~al.}(2014){Huber}, {Silva Aguirre}, {Matthews},
  {Pinsonneault}, {Gaidos}, {Garc{\'{\i}}a}, {Hekker}, {Mathur}, {Mosser},
  {Torres}, {Bastien}, {Basu}, {Bedding}, {Chaplin}, {Demory}, {Fleming},
  {Guo}, {Mann}, {Rowe}, {Serenelli}, {Smith}, \&
  {Stello}}]{2014ApJS..211....2H}
{Huber}, D., {Silva Aguirre}, V., {Matthews}, J.~M., {et~al.} 2014, \apjs, 211,
  2

\bibitem[{{Huber} {et~al.}(2017){Huber}, {Zinn}, {Bojsen-Hansen},
  {Pinsonneault}, {Sahlholdt}, {Serenelli}, {Silva Aguirre}, {Stassun},
  {Stello}, {Tayar}, {Bastien}, {Bedding}, {Buchhave}, {Chaplin}, {Davies},
  {Garc{\'{\i}}a}, {Latham}, {Mathur}, {Mosser}, \&
  {Sharma}}]{2017ApJ...844..102H}
{Huber}, D., {Zinn}, J., {Bojsen-Hansen}, M., {et~al.} 2017, \apj, 844, 102

\bibitem[{{Jenkins} {et~al.}(2015){Jenkins}, {Twicken}, {Batalha}, {Caldwell},
  {Cochran}, {Endl}, {Latham}, {Esquerdo}, {Seader}, {Bieryla}, {Petigura},
  {Ciardi}, {Marcy}, {Isaacson}, {Huber}, {Rowe}, {Torres}, {Bryson},
  {Buchhave}, {Ramirez}, {Wolfgang}, {Li}, {Campbell}, {Tenenbaum},
  {Sanderfer}, {Henze}, {Catanzarite}, {Gilliland}, \&
  {Borucki}}]{2015AJ....150...56J}
{Jenkins}, J.~M., {Twicken}, J.~D., {Batalha}, N.~M., {et~al.} 2015, \aj, 150,
  56

\bibitem[{{Kamiaka} {et~al.}(2018){Kamiaka}, {Benomar}, \&
  {Suto}}]{2018MNRAS.479..391K}
{Kamiaka}, S., {Benomar}, O., \& {Suto}, Y. 2018, \mnras, 479, 391

\bibitem[{{Kjeldsen} \& {Bedding}(1995)}]{1995A&A...293...87K}
{Kjeldsen}, H. \& {Bedding}, T.~R. 1995, \aap, 293, 87

\bibitem[{{K{\"o}nigl} {et~al.}(2017){K{\"o}nigl}, {Giacalone}, \&
  {Matsakos}}]{2017ApJ...846L..13K}
{K{\"o}nigl}, A., {Giacalone}, S., \& {Matsakos}, T. 2017, \apjl, 846, L13

\bibitem[{{Lai} {et~al.}(2011){Lai}, {Foucart}, \& {Lin}}]{2011MNRAS.412.2790L}
{Lai}, D., {Foucart}, F., \& {Lin}, D.~N.~C. 2011, \mnras, 412, 2790

\bibitem[{{Lebreton} \& {Goupil}(2014)}]{2014A&A...569A..21L}
{Lebreton}, Y. \& {Goupil}, M.~J. 2014, \aap, 569, A21

\bibitem[{{Li} \& {Winn}(2016)}]{2016ApJ...818....5L}
{Li}, G. \& {Winn}, J.~N. 2016, \apj, 818, 5

\bibitem[{{Lin} {et~al.}(1996){Lin}, {Bodenheimer}, \&
  {Richardson}}]{1996Natur.380..606L}
{Lin}, D.~N.~C., {Bodenheimer}, P., \& {Richardson}, D.~C. 1996, \nat, 380, 606

\bibitem[{{Lund} {et~al.}(2017){Lund}, {Silva Aguirre}, {Davies}, {Chaplin},
  {Christensen-Dalsgaard}, {Houdek}, {White}, {Bedding}, {Ball}, {Huber},
  {Antia}, {Lebreton}, {Latham}, {Handberg}, {Verma}, {Basu}, {Casagrande},
  {Justesen}, {Kjeldsen}, \& {Mosumgaard}}]{2017ApJ...835..172L}
{Lund}, M.~N., {Silva Aguirre}, V., {Davies}, G.~R., {et~al.} 2017, \apj, 835,
  172

\bibitem[{{Lundkvist} {et~al.}(2018){Lundkvist}, {Huber}, {Silva Aguirre}, \&
  {Chaplin}}]{2018arXiv180402214L}
{Lundkvist}, M.~S., {Huber}, D., {Silva Aguirre}, V., \& {Chaplin}, W.~J. 2018,
  in {Handbook of Exoplanets}, ed. {{Deeg}, Hans J. and {Belmonte}, Juan
  Antonio} ({Springer}), 24

\bibitem[{{Marcy} {et~al.}(2014){Marcy}, {Isaacson}, {Howard}, {Rowe},
  {Jenkins}, {Bryson}, {Latham}, {Howell}, {Gautier}, {Batalha}, {Rogers},
  {Ciardi}, {Fischer}, {Gilliland}, {Kjeldsen}, {Christensen-Dalsgaard},
  {Huber}, {Chaplin}, {Basu}, {Buchhave}, {Quinn}, {Borucki}, {Koch}, {Hunter},
  {Caldwell}, {Van Cleve}, {Kolbl}, {Weiss}, {Petigura}, {Seager}, {Morton},
  {Johnson}, {Ballard}, {Burke}, {Cochran}, {Endl}, {MacQueen}, {Everett},
  {Lissauer}, {Ford}, {Torres}, {Fressin}, {Brown}, {Steffen}, {Charbonneau},
  {Basri}, {Sasselov}, {Winn}, {Sanchis-Ojeda}, {Christiansen}, {Adams},
  {Henze}, {Dupree}, {Fabrycky}, {Fortney}, {Tarter}, {Holman}, {Tenenbaum},
  {Shporer}, {Lucas}, {Welsh}, {Orosz}, {Bedding}, {Campante}, {Davies},
  {Elsworth}, {Handberg}, {Hekker}, {Karoff}, {Kawaler}, {Lund}, {Lundkvist},
  {Metcalfe}, {Miglio}, {Silva Aguirre}, {Stello}, {White}, {Boss}, {Devore},
  {Gould}, {Prsa}, {Agol}, {Barclay}, {Coughlin}, {Brugamyer}, {Mullally},
  {Quintana}, {Still}, {Thompson}, {Morrison}, {Twicken}, {D{\'e}sert},
  {Carter}, {Crepp}, {H{\'e}brard}, {Santerne}, {Moutou}, {Sobeck}, {Hudgins},
  {Haas}, {Robertson}, {Lillo-Box}, \& {Barrado}}]{2014ApJS..210...20M}
{Marcy}, G.~W., {Isaacson}, H., {Howard}, A.~W., {et~al.} 2014, \apjs, 210, 20

\bibitem[{{Marrese} {et~al.}(2018){Marrese}, {Marinoni}, {Fabrizio}, \&
  {Altavilla}}]{2018arXiv180809151M}
{Marrese}, P.~M., {Marinoni}, S., {Fabrizio}, M., \& {Altavilla}, G. 2018,
  ArXiv e-prints [\eprint[arXiv]{1808.09151}]

\bibitem[{{Massey} \& {Hanson}(2013)}]{2013pss2.book...35M}
{Massey}, P. \& {Hanson}, M.~M. 2013, {Astronomical Spectroscopy} (Springer),
  35

\bibitem[{{Mathur} {et~al.}(2012){Mathur}, {Metcalfe}, {Woitaszek}, {Bruntt},
  {Verner}, {Christensen-Dalsgaard}, {Creevey}, {Do{\v g}an}, {Basu}, {Karoff},
  {Stello}, {Appourchaux}, {Campante}, {Chaplin}, {Garc{\'{\i}}a}, {Bedding},
  {Benomar}, {Bonanno}, {Deheuvels}, {Elsworth}, {Gaulme}, {Guzik}, {Handberg},
  {Hekker}, {Herzberg}, {Monteiro}, {Piau}, {Quirion}, {R{\'e}gulo}, {Roth},
  {Salabert}, {Serenelli}, {Thompson}, {Trampedach}, {White}, {Ballot},
  {Brand{\~a}o}, {Molenda-{\.Z}akowicz}, {Kjeldsen}, {Twicken}, {Uddin}, \&
  {Wohler}}]{2012ApJ...749..152M}
{Mathur}, S., {Metcalfe}, T.~S., {Woitaszek}, M., {et~al.} 2012, \apj, 749, 152

\bibitem[{{Matsakos} \& {K{\"o}nigl}(2015)}]{2015ApJ...809L..20M}
{Matsakos}, T. \& {K{\"o}nigl}, A. 2015, \apjl, 809, L20

\bibitem[{{Matsakos} \& {K{\"o}nigl}(2017)}]{2017AJ....153...60M}
{Matsakos}, T. \& {K{\"o}nigl}, A. 2017, \aj, 153, 60

\bibitem[{{Metcalfe} {et~al.}(2014){Metcalfe}, {Creevey}, {Do{\u g}an},
  {Mathur}, {Xu}, {Bedding}, {Chaplin}, {Christensen-Dalsgaard}, {Karoff},
  {Trampedach}, {Benomar}, {Brown}, {Buzasi}, {Campante}, {{\c C}elik},
  {Cunha}, {Davies}, {Deheuvels}, {Derekas}, {Di Mauro}, {Garc{\'{\i}}a},
  {Guzik}, {Howe}, {MacGregor}, {Mazumdar}, {Montalb{\'a}n}, {Monteiro},
  {Salabert}, {Serenelli}, {Stello}, {Ste{\c s}acute}, {licki}, {Suran},
  {Y{\i}ld{\i}z}, {Aksoy}, {Elsworth}, {Gruberbauer}, {Guenther}, {Lebreton},
  {Molaverdikhani}, {Pricopi}, {Simoniello}, \& {White}}]{2014ApJS..214...27M}
{Metcalfe}, T.~S., {Creevey}, O.~L., {Do{\u g}an}, G., {et~al.} 2014, \apjs,
  214, 27

\bibitem[{{Morton} \& {Johnson}(2011)}]{2011ApJ...729..138M}
{Morton}, T.~D. \& {Johnson}, J.~A. 2011, \apj, 729, 138

\bibitem[{{Mosser} {et~al.}(2013){Mosser}, {Michel}, {Belkacem}, {Goupil},
  {Baglin}, {Barban}, {Provost}, {Samadi}, {Auvergne}, \&
  {Catala}}]{2013A&A...550A.126M}
{Mosser}, B., {Michel}, E., {Belkacem}, K., {et~al.} 2013, \aap, 550, A126

\bibitem[{{Nissen} {et~al.}(2017){Nissen}, {Silva Aguirre},
  {Christensen-Dalsgaard}, {Collet}, {Grundahl}, \&
  {Slumstrup}}]{2017A&A...608A.112N}
{Nissen}, P.~E., {Silva Aguirre}, V., {Christensen-Dalsgaard}, J., {et~al.}
  2017, \aap, 608, A112

\bibitem[{{P{\'a}l} {et~al.}(2008){P{\'a}l}, {Bakos}, {Torres}, {Noyes},
  {Latham}, {Kov{\'a}cs}, {Marcy}, {Fischer}, {Butler}, {Sasselov}, {Sip{\H
  o}cz}, {Esquerdo}, {Kov{\'a}cs}, {Stefanik}, {L{\'a}z{\'a}r}, {Papp}, \&
  {S{\'a}ri}}]{2008ApJ...680.1450P}
{P{\'a}l}, A., {Bakos}, G.~{\'A}., {Torres}, G., {et~al.} 2008, \apj, 680, 1450

\bibitem[{{Paxton} {et~al.}(2011){Paxton}, {Bildsten}, {Dotter}, {Herwig},
  {Lesaffre}, \& {Timmes}}]{2011ApJS..192....3P}
{Paxton}, B., {Bildsten}, L., {Dotter}, A., {et~al.} 2011, \apjs, 192, 3

\bibitem[{{Paxton} {et~al.}(2013){Paxton}, {Cantiello}, {Arras}, {Bildsten},
  {Brown}, {Dotter}, {Mankovich}, {Montgomery}, {Stello}, {Timmes}, \&
  {Townsend}}]{2013ApJS..208....4P}
{Paxton}, B., {Cantiello}, M., {Arras}, P., {et~al.} 2013, \apjs, 208, 4

\bibitem[{{Paxton} {et~al.}(2015){Paxton}, {Marchant}, {Schwab}, {Bauer},
  {Bildsten}, {Cantiello}, {Dessart}, {Farmer}, {Hu}, {Langer}, {Townsend},
  {Townsley}, \& {Timmes}}]{2015ApJS..220...15P}
{Paxton}, B., {Marchant}, P., {Schwab}, J., {et~al.} 2015, \apjs, 220, 15

\bibitem[{{Paxton} {et~al.}(2018){Paxton}, {Schwab}, {Bauer}, {Bildsten},
  {Blinnikov}, {Duffell}, {Farmer}, {Goldberg}, {Marchant}, {Sorokina},
  {Thoul}, {Townsend}, \& {Timmes}}]{2018ApJS..234...34P}
{Paxton}, B., {Schwab}, J., {Bauer}, E.~B., {et~al.} 2018, \apjs, 234, 34

\bibitem[{{Plavchan} \& {Bilinski}(2013)}]{2013ApJ...769...86P}
{Plavchan}, P. \& {Bilinski}, C. 2013, \apj, 769, 86

\bibitem[{{Pr{\v s}a} {et~al.}(2016){Pr{\v s}a}, {Harmanec}, {Torres},
  {Mamajek}, {Asplund}, {Capitaine}, {Christensen-Dalsgaard}, {Depagne},
  {Haberreiter}, {Hekker}, {Hilton}, {Kopp}, {Kostov}, {Kurtz}, {Laskar},
  {Mason}, {Milone}, {Montgomery}, {Richards}, {Schmutz}, {Schou}, \&
  {Stewart}}]{2016AJ....152...41P}
{Pr{\v s}a}, A., {Harmanec}, P., {Torres}, G., {et~al.} 2016, \aj, 152, 41

\bibitem[{{Rogers}(2015)}]{2015ApJ...801...41R}
{Rogers}, L.~A. 2015, \apj, 801, 41

\bibitem[{{Rowe} {et~al.}(2014){Rowe}, {Bryson}, {Marcy}, {Lissauer},
  {Jontof-Hutter}, {Mullally}, {Gilliland}, {Issacson}, {Ford}, {Howell},
  {Borucki}, {Haas}, {Huber}, {Steffen}, {Thompson}, {Quintana}, {Barclay},
  {Still}, {Fortney}, {Gautier}, {Hunter}, {Caldwell}, {Ciardi}, {Devore},
  {Cochran}, {Jenkins}, {Agol}, {Carter}, \& {Geary}}]{2014ApJ...784...45R}
{Rowe}, J.~F., {Bryson}, S.~T., {Marcy}, G.~W., {et~al.} 2014, \apj, 784, 45

\bibitem[{{Roxburgh} \& {Vorontsov}(2003)}]{2003A&A...411..215R}
{Roxburgh}, I.~W. \& {Vorontsov}, S.~V. 2003, \aap, 411, 215

\bibitem[{{Safonova} {et~al.}(2016){Safonova}, {Murthy}, \&
  {Shchekinov}}]{2016IJAsB..15...93S}
{Safonova}, M., {Murthy}, J., \& {Shchekinov}, Y.~A. 2016, International
  Journal of Astrobiology, 15, 93

\bibitem[{{Sahlholdt} \& {Silva Aguirre}(2018)}]{2018MNRAS.481L.125S}
{Sahlholdt}, C.~L. \& {Silva Aguirre}, V. 2018, \mnras, 481, L125

\bibitem[{{Seager} {et~al.}(2007){Seager}, {Kuchner}, {Hier-Majumder}, \&
  {Militzer}}]{2007ApJ...669.1279S}
{Seager}, S., {Kuchner}, M., {Hier-Majumder}, C.~A., \& {Militzer}, B. 2007,
  \apj, 669, 1279

\bibitem[{{Seager} \& {Mall{\'e}n-Ornelas}(2003)}]{2003ApJ...585.1038S}
{Seager}, S. \& {Mall{\'e}n-Ornelas}, G. 2003, \apj, 585, 1038

\bibitem[{{Sharma} {et~al.}(2016){Sharma}, {Stello}, {Bland-Hawthorn}, {Huber},
  \& {Bedding}}]{2016ApJ...822...15S}
{Sharma}, S., {Stello}, D., {Bland-Hawthorn}, J., {Huber}, D., \& {Bedding},
  T.~R. 2016, \apj, 822, 15

\bibitem[{{Silva Aguirre} {et~al.}(2018){Silva Aguirre}, {Bojsen-Hansen},
  {Slumstrup}, {Casagrande}, {Kawata}, {Ciuc{\v a}}, {Handberg}, {Lund},
  {Mosumgaard}, {Huber}, {Johnson}, {Pinsonneault}, {Serenelli}, {Stello},
  {Tayar}, {Bird}, {Cassisi}, {Hon}, {Martig}, {Nissen}, {Rix},
  {Sch{\"o}nrich}, {Sahlholdt}, {Trick}, \& {Yu}}]{2018MNRAS.475.5487S}
{Silva Aguirre}, V., {Bojsen-Hansen}, M., {Slumstrup}, D., {et~al.} 2018,
  \mnras, 475, 5487

\bibitem[{{Silva Aguirre} {et~al.}(2015){Silva Aguirre}, {Davies}, {Basu},
  {Christensen-Dalsgaard}, {Creevey}, {Metcalfe}, {Bedding}, {Casagrande},
  {Handberg}, {Lund}, {Nissen}, {Chaplin}, {Huber}, {Serenelli}, {Stello}, {Van
  Eylen}, {Campante}, {Elsworth}, {Gilliland}, {Hekker}, {Karoff}, {Kawaler},
  {Kjeldsen}, \& {Lundkvist}}]{2015MNRAS.452.2127S}
{Silva Aguirre}, V., {Davies}, G.~R., {Basu}, S., {et~al.} 2015, \mnras, 452,
  2127

\bibitem[{{Silva Aguirre} {et~al.}(2017){Silva Aguirre}, {Lund}, {Antia},
  {Ball}, {Basu}, {Christensen-Dalsgaard}, {Lebreton}, {Reese}, {Verma},
  {Casagrande}, {Justesen}, {Mosumgaard}, {Chaplin}, {Bedding}, {Davies},
  {Handberg}, {Houdek}, {Huber}, {Kjeldsen}, {Latham}, {White}, {Coelho},
  {Miglio}, \& {Rendle}}]{2017ApJ...835..173S}
{Silva Aguirre}, V., {Lund}, M.~N., {Antia}, H.~M., {et~al.} 2017, \apj, 835,
  173

\bibitem[{{Steffen} {et~al.}(2013){Steffen}, {Fabrycky}, {Agol}, {Ford},
  {Morehead}, {Cochran}, {Lissauer}, {Adams}, {Borucki}, {Bryson}, {Caldwell},
  {Dupree}, {Jenkins}, {Robertson}, {Rowe}, {Seader}, {Thompson}, \&
  {Twicken}}]{2013MNRAS.428.1077S}
{Steffen}, J.~H., {Fabrycky}, D.~C., {Agol}, E., {et~al.} 2013, \mnras, 428,
  1077

\bibitem[{{Steffen} {et~al.}(2012){Steffen}, {Fabrycky}, {Ford}, {Carter},
  {D{\'e}sert}, {Fressin}, {Holman}, {Lissauer}, {Moorhead}, {Rowe},
  {Ragozzine}, {Welsh}, {Batalha}, {Borucki}, {Buchhave}, {Bryson}, {Caldwell},
  {Charbonneau}, {Ciardi}, {Cochran}, {Endl}, {Everett}, {Gautier},
  {Gilliland}, {Girouard}, {Jenkins}, {Horch}, {Howell}, {Isaacson}, {Klaus},
  {Koch}, {Latham}, {Li}, {Lucas}, {MacQueen}, {Marcy}, {McCauliff}, {Middour},
  {Morris}, {Mullally}, {Quinn}, {Quintana}, {Shporer}, {Still}, {Tenenbaum},
  {Thompson}, {Twicken}, \& {Van Cleve}}]{2012MNRAS.421.2342S}
{Steffen}, J.~H., {Fabrycky}, D.~C., {Ford}, E.~B., {et~al.} 2012, \mnras, 421,
  2342

\bibitem[{{Teyssandier} {et~al.}(2013){Teyssandier}, {Terquem}, \&
  {Papaloizou}}]{2013MNRAS.428..658T}
{Teyssandier}, J., {Terquem}, C., \& {Papaloizou}, J.~C.~B. 2013, \mnras, 428,
  658

\bibitem[{{Theme{\ss}l} {et~al.}(2018){Theme{\ss}l}, {Hekker}, {Southworth},
  {Beck}, {Pavlovski}, {Tkachenko}, {Angelou}, {Ball}, {Barban}, {Corsaro},
  {Elsworth}, {Handberg}, \& {Kallinger}}]{2018MNRAS.478.4669T}
{Theme{\ss}l}, N., {Hekker}, S., {Southworth}, J., {et~al.} 2018, \mnras, 478,
  4669

\bibitem[{{Thies} {et~al.}(2011){Thies}, {Kroupa}, {Goodwin}, {Stamatellos}, \&
  {Whitworth}}]{2011MNRAS.417.1817T}
{Thies}, I., {Kroupa}, P., {Goodwin}, S.~P., {Stamatellos}, D., \& {Whitworth},
  A.~P. 2011, \mnras, 417, 1817

\bibitem[{{Torres} {et~al.}(2010){Torres}, {Andersen}, \&
  {Gim{\'e}nez}}]{2010A&ARv..18...67T}
{Torres}, G., {Andersen}, J., \& {Gim{\'e}nez}, A. 2010, \aapr, 18, 67

\bibitem[{{Townsend} \& {Teitler}(2013)}]{2013MNRAS.435.3406T}
{Townsend}, R.~H.~D. \& {Teitler}, S.~A. 2013, \mnras, 435, 3406

\bibitem[{{Ulrich}(1986)}]{1986ApJ...306L..37U}
{Ulrich}, R.~K. 1986, \apjl, 306, L37

\bibitem[{{{\v S}koda}(2017)}]{2011arXiv1112.2787S}
{{\v S}koda}, P. 2017, in Proceedings of EURO-VO Workshop Astronomical
  Spectroscopy and Virtual Observatory

\bibitem[{{Valle} {et~al.}(2018){Valle}, {Dell'Omodarme}, {Prada Moroni}, \&
  {Degl'Innocenti}}]{2018arXiv181006997V}
{Valle}, G., {Dell'Omodarme}, M., {Prada Moroni}, P.~G., \& {Degl'Innocenti},
  S. 2018, ArXiv e-prints [\eprint[arXiv]{1810.06997}]

\bibitem[{{Van Eylen} \& {Albrecht}(2015)}]{2015ApJ...808..126V}
{Van Eylen}, V. \& {Albrecht}, S. 2015, \apj, 808, 126

\bibitem[{{Van Eylen} {et~al.}(2014){Van Eylen}, {Lund}, {Silva Aguirre},
  {Arentoft}, {Kjeldsen}, {Albrecht}, {Chaplin}, {Isaacson}, {Pedersen},
  {Jessen-Hansen}, {Tingley}, {Christensen-Dalsgaard}, {Aerts}, {Campante}, \&
  {Bryson}}]{2014ApJ...782...14V}
{Van Eylen}, V., {Lund}, M.~N., {Silva Aguirre}, V., {et~al.} 2014, \apj, 782,
  14

\bibitem[{{Viani} {et~al.}(2017){Viani}, {Basu}, {Chaplin}, {Davies}, \&
  {Elsworth}}]{2017ApJ...843...11V}
{Viani}, L.~S., {Basu}, S., {Chaplin}, W.~J., {Davies}, G.~R., \& {Elsworth},
  Y. 2017, \apj, 843, 11

\bibitem[{{Watson} {et~al.}(2011){Watson}, {Littlefair}, {Diamond}, {Collier
  Cameron}, {Fitzsimmons}, {Simpson}, {Moulds}, \&
  {Pollacco}}]{2011MNRAS.413L..71W}
{Watson}, C.~A., {Littlefair}, S.~P., {Diamond}, C., {et~al.} 2011, \mnras,
  413, L71

\bibitem[{{Weiss} \& {Marcy}(2014)}]{2014ApJ...783L...6W}
{Weiss}, L.~M. \& {Marcy}, G.~W. 2014, \apjl, 783, L6

\bibitem[{{White} {et~al.}(2011){White}, {Bedding}, {Stello},
  {Christensen-Dalsgaard}, {Huber}, \& {Kjeldsen}}]{2011ApJ...743..161W}
{White}, T.~R., {Bedding}, T.~R., {Stello}, D., {et~al.} 2011, \apj, 743, 161

\bibitem[{{Xie}(2014)}]{2014ApJS..210...25X}
{Xie}, J.-W. 2014, \apjs, 210, 25

\end{thebibliography}
\end{multicols}
\vspace*{0.1\baselineskip}
\begin{acknowledgements}
    We thank Didier Queloz and Carla Wiles for useful discussions. 
    We thank the anonymous referee for their feedback, which improved the manuscript. 
    We also thank the developers of the {\tt astroquery} package \citep{adam_ginsburg_2018_1160627} and the NASA Exoplanet Archive \citep{2013PASP..125..989A}. 
    Funding for the Stellar Astrophysics Centre is provided by The Danish National Research Foundation (Grant agreement no.: DNRF106). 
    The research leading to the presented results has received funding from the European Research Council under the European Community's Seventh Framework Programme (FP7/2007-2013) / ERC grant agreement no 338251 (StellarAges). 
    S.B.\ acknowledges partial support from NSF grant AST-1514676 and NASA grant NNX13AE70G.
\end{acknowledgements}
%\end{multicols}
%\vfill{}
\end{document}